\documentclass[]{aa}

\usepackage{graphicx}
\usepackage{txfonts}
\usepackage{lscape}
\usepackage{placeins}
\usepackage{rotating}
\usepackage{tikz}
\usepackage{pdflscape}
\usepackage{xcolor}
\usepackage[colorlinks=true,citecolor=blue,linkcolor=blue,urlcolor=blue]{hyperref}
\begin{document}

\title{
  {\it JWST} detection of extremely excited outflowing CO and
  H$_2$O in~VV~114~E~SW: a possible rapidly accreting IMBH}

   \author{Eduardo Gonz\'alez-Alfonso
          \inst{1}
          \and
          Ismael Garc\'{\i}a-Bernete
          \inst{2}
          \and
          Miguel Pereira-Santaella
          \inst{3}
          \and
          David A. Neufeld
          \inst{4}
          \and
          Jacqueline Fischer
          \inst{5}
           \and
          Fergus R. Donnan
          \inst{2}
         }

   \institute{Universidad de Alcal\'a, Departamento de F\'{\i}sica
     y Matem\'aticas, Campus Universitario, E-28871 Alcal\'a de Henares,
     Madrid, Spain\\
              \email{eduardo.gonzalez@uah.es}
              \and
   Department of Physics, University of Oxford, Keble Road,
                Oxford OX1 3RH, UK         
              \and
   Instituto de F\'{\i}sica Fundamental, CSIC, Calle Serrano 123,
                28006 Madrid, Spain         
              \and
              William H. Miller Department of Physics \& Astronomy, Johns
              Hopkins University, Baltimore, MD 21218, USA            
              \and
              George Mason University, Department of Physics \& Astronomy,
              MS 3F3, 4400 University Drive, Fairfax, VA 22030, USA
             }

   \authorrunning{Gonz\'alez-Alfonso et al.}
   \titlerunning{A possible rapidly accreting IMBH in VV 114 E SW}
   

 
   \abstract{Mid-infrared (mid-IR) gas-phase molecular bands are powerful
     diagnostics of the warm interstellar medium.
     We report the {\it James Webb Space Telescope}
     detection of the CO $v=1-0$ ($4.4-5.0$\,$\mu$m) and H$_2$O
     $\nu_2=1-0$ ($5.0-7.8$\,$\mu$m) ro-vibrational bands, 
     both in absorption, toward the
     ``s2'' core in the southwest nucleus of the
       merging galaxy VV 114 E. All ro-vibrational CO lines 
     up to $J_{\mathrm{low}}=33$ ($E_{\mathrm{low}}\approx3000$\,K) are detected,
     as well as a forest of H$_2$O lines up to
     $13_{0,13}$ ($E_{\mathrm{low}}\approx2600$\,K).
     The highest-excitation lines are blueshifted by
     $\sim180$\,km\,s$^{-1}$ relative
     to the extended molecular cloud, which is traced by the 
     rotational CO\,($J=3-2$) 346\,GHz line observed with
     the Atacama Large Millimeter/submillimeter Array.
     The bands also show absorption in a low-velocity
     component (blueshifted by $\approx30$\,km\,s$^{-1}$) with lower
     excitation. The analysis shows that the bands are observed against
     a continuum with effective temperature of $T_{\mathrm{bck}}\sim550$\,K
     extinguished with     
     $\tau_{\mathrm{6\mu m}}^{\mathrm{ext}}\sim 2.5-3$
     ($A_k\sim6.9-8.3$\,mag).
     The high-excitation CO and H$_2$O lines are consistent with $v=0$
     thermalization with $T_{\mathrm{rot}}\approx450$\,K and column densities of
     $N_{\mathrm{CO}}\approx(1.7-3.5)\times10^{19}$\,cm$^{-2}$ and
     $N_{\mathrm{H_2O}}\approx(1.5-3.0)\times10^{19}$\,cm$^{-2}$.
     Thermalization of the $v=0$ levels of H$_2$O
      requires either an extreme density of
     $n_{\mathrm{H_2}}\gtrsim10^9$\,cm$^{-3}$, or radiative excitation
     by the mid-IR field in a very compact ($<1$\,pc) optically thick source
     emitting $\sim10^{10}\,L_{\odot}$.
     The latter alternative is favored, implying that
     the observed absorption 
     probes the very early stages of a fully enshrouded
     active black hole (BH). On the basis of a simple model for
       BH growth and applying a lifetime constraint to the s2 core,
       an intermediate-mass BH
       (IMBH, $M_{\mathrm{BH}}\sim4.5\times10^4\,M_{\odot}$)
     accreting at super-Eddington rates is suggested,
     where the observed feedback has not yet been able to 
     break through the natal cocoon.
}

   \keywords{Galaxies: evolution  --
               Galaxies: nuclei  --
               Infrared: galaxies  --
               }

   \maketitle

\section{Introduction}
\label{intro}

Major galaxy mergers leading to (ultra-)luminous infrared
  galaxies ((U)LIRGs) are considered important
  precursors for the formation and growth of super massive black holes
  (SMBHs, $>10^6\,M_{\odot}$) in the local universe \citep{san96}.
  While the general process,
  involving tidal forces that efficiently funnel gas towards
  the galactic center(s) \citep[e.g.,][]{hop06}, is well constrained
  observationally, the earliest
  phases of intermediate-mass black hole (IMBH, $\sim10^{2-5}\,M_{\odot}$)
  formation and growth in mergers are unconstrained
  \citep[see recent reviews by][]{gre20,ask23}.
  Close to the high end of the IMBH mass range, the (so far) few
  identifications have been based on dynamical measurements in
  nearby, mostly dwarf galaxies
  \citep[$\lesssim10$\,Mpc; e.g.,][]{ngu19,her05,pec22}, and beyond the
  Local Group from optical spectroscopy
  \citep[i.e., the broad and narrow line regions,][]{bal15},
  X-ray emission \citep{dav11}, and tidal disruption events
  \citep[TDEs,][]{lin18}. At the lower end, detections come from
  gravitational wave events \citep{abb20}.
  Potential tracers are also radio emission,
  mid-infrared (mid-IR) colors, polycyclic aromatic hydrocarbons
  (PAHs), and spectroscopy of high ionization 
  potential ions \citep[e.g.,][and references therein]{gre20}.
  However, some of these measurements are challenging due to low-level
  emission and potential confusion with stellar activity
  \citep{cas15,gre20,ask23}.
  
  To date, there have been no reported detections of IMBHs in (U)LIRG mergers.
  In wet mergers, some of the above diagnostics
  are critical at early stages of IMBH growth when
  high amounts of gas and dust may shrink the highly ionized regions
  \citep{toyi21} and obscure the most well known tracers of active galactic
  nuclei (AGN). Nevertheless, with large reservoirs of gas available, an 
  IMBH may be accreting at super- or even hyper-Eddington rates such that
  mid-IR observations can trace the surrounding hot dust
  \citep[e.g.,][]{ina16}.
  To identify these high concentrations of hot dust, 
  mid-IR spectroscopy of absorption bands of gas-phase molecular species,
  namely CO and H$_2$O, are essential probes.

The gas-phase CO $v=1-0$ 4.7\,$\mu$m and H$_2$O $\nu_2=1-0$ 6.3\,$\mu$m
fundamental bands are powerful mid-IR diagnostics of the
interestellar medium (ISM):
they probe the physical conditions and gas kinematics by sampling all rotational
levels of the ground vibrational state that are significantly populated,
and trace the chemistry of the gas via the
[CO]/[H$_2$O] abundance ratio.
Since the bands are excited by the mid-IR radiation field, they can also give
clues on the ISM geometry relative to the
mid-IR luminosity sources in the region,
depending on whether the bands are detected in emission, in absorption,
or some lines in absorption (within the R-branch) and some in emission
(the P-branch). In addition,
the relative strength of the ro-vibrational lines can potentially give clues on
the gas column density, and on the slope of the exciting continuum
emitted by hot dust --and thus on its effective temperature.
Previous studies of the CO band in extragalactic sources
  have been carried out with high spectral resolution
  using ground-based facilities \citep{geb06,shi13,oni21},
  and with low spectral resolution using {\it Spitzer} and {\it AKARI}
  \citep{spo04,bab18}.
The unprecedented sensitivity of
the {\it James Webb Space Telescope (JWST)}
with the high spectral and spatial resolution
provided by the Near-Infrared Spectrograph \citep[NIRSpec,][]{jak22,bok22}
and the Mid-Infrared Instrument/Medium-resolution spectroscopy
\citep[MIRI/MRS,][]{rie15,wri15}
is ideal to further extend
the study of the molecular bands in galaxies
\citep{per23,gber23}.
We take all spectroscopic parameters for CO and H$_2$O
from the HITRAN2020 database \citep{gor22}.

\begin{figure*}[h]
   \centering
\includegraphics[width=17.0cm]{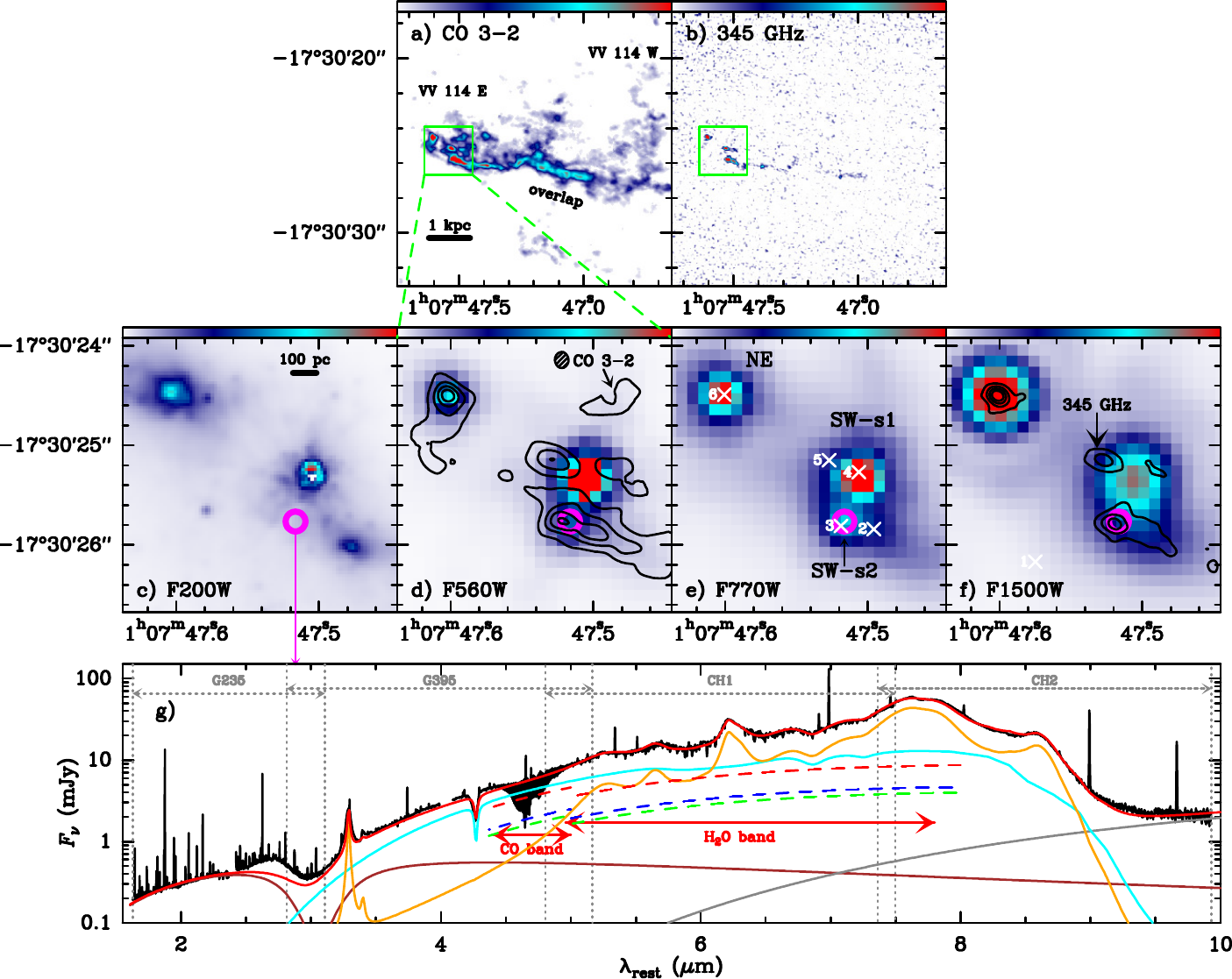}
\caption[]{{\bf a} and {\bf b}: ALMA maps of CO\,($3-2$) (moment 0)
  and the 345 GHz continuum of the VV 114 system as observed with ALMA.
  The green box in both panels indicates the nuclear region of VV 114 E
  mapped in the middle panels. {\bf c-f}: {\it JWST} 
  NIRCam F200W, MIRI F560W, F770W, and F1500 images of the VV 114 E nucleus.
  Overlaid in panels d and f are the contours of CO\,($3-2$) and 345\,GHz
  continuum, respectively. The ALMA beam ($0.16''\times0.14''$)
  is shown in panel d.
  The 33\,GHz continuum sources \citep{son22} are indicated with
  crosses and labelled in panel e, where we also label the main mid-IR peaks
  (NE, SW-s1, and SW-s2). The magenta circle, coincident with SW-s2
  and 33\,GHz source \#3, indicates the position where the extreme excitation
  of CO and H$_2$O is found. {\bf g}: The near- and mid-IR spectrum extracted
  from SW-s2. The extent of the CO and H$_2$O bands,
  and of the NIRSpec gratings and MIRI/MRS channels, are indicated.
  The solid curves show a model for the continuum, including
  the PAHs (orange) and three blackbody sources with temperatures
  1360 (brown), 550 (light-blue), and 230\,K (gray); red is total.
  The light-blue component, dominating the continuum associated with the
    molecular bands, is attenuated by foreground material with
    $\tau_{\mathrm{6\mu m}}^{\mathrm{ext}}=2.5$ (see text). 
    The dashed lines indicate the
  continuum covered by CO and H$_2$O in the hot ($H_C$, in blue), 
  the warm ($W_C$, in green), and the sum of both (in red),
  as predicted by the fiducial model for the bands. 
}
\label{maps}%
\end{figure*}

Here we report and analyze the detection of extremely excited CO and H$_2$O,
observed in absorption, in the local merger VV 114
(IC\,1623, IRAS\,F01053$-$1746), a luminous infrared (LIRG,
$L_{\mathrm{IR}}=5\times10^{11}$\,$L_{\odot}$) mid-stage merger \citep{arm09} with
the two galaxy components separated by $\sim6$\,kpc.
While the western galaxy is bright in the UV
\citep{gol02} and optical \citep{kno94} showing low visual extinction, a large
concentration of dust obscures the eastern component (VV 114 E), which
dominates the mid-IR emission. VV 114 E has been imaged in several
molecular lines at millimeter wavelengths \citep[e.g.][]{ion13,sai15,sai17},
showing enhanced emission along a narrow, 4\,kpc-long dense filament in the
east-west direction (including an overlap region between the two galaxies,
see also Fig.~\ref{maps}a,b), in which Pa$\alpha$ emission
  indicates ongoing star formation \citep{tat12,ion13}.
This filament, which is resolved into dense clumps of
several\,$\times10^6$\,$M_{\odot}$, suggests widespread shocks triggered by
the dynamic interaction of the merging disk galaxies.
The nucleus of VV 114 E is located at the easternmost extreme edge of the
filament, and is also resolved into several massive clumps. {\it JWST}/MIRI
photometry of up to 40 star-forming knots has been reported by \cite{eva22}.
The possible presence of an AGN in VV 114 E has been debated for years.
Analysis of X-ray emission from VV 114 E has not yielded
  conclusive results on the nature of the source, which has a spectrum harder
  than other Chandra-detected point sources in the galaxy \citep{gri06,gar20}.
\cite{sai15} favor an AGN at the NE core based on the HCN/HCO$^+$
ratio. Based on the PAH emission, \cite{don23} identified a
deeply obscured nucleus at NE that could extinguish
the undetected coronal lines.
On the other hand, \cite{eva22} and \cite{ric23}
proposed that the AGN is located at the SW-s1 clump (see Fig~\ref{maps}e)
based on the low PAH equivalent widths and mid- and near-IR colors.
To complicate things, the extreme source of CO and H$_2$O
excitation reported here is however located at the SW-s2 knot. We adopt a
distance to VV 114 of 88\,Mpc.

\section{Observations and results}
\label{obs}

We have used the MIRI (imaging and MRS), NIRCam, and NIRSpec IFU data
on VV 114 from the
Director’s Discretionary Early Release
Science (DD-ERS) Program \#1328 (PI: L.~Armus and A.~Evans).
The observations and data reduction are
described in Appendix~\ref{reduc}.
The images of the VV 114 E nucleus in the NIRCam F200W
($1.7-2.2$\,$\mu$m), and MIRI
F560W, F770W, and F1500W
($5.0-6.2$, $6.6-8.7$, and $13.5-16.6$\,$\mu$m)
filters are shown 
in Fig.~\ref{maps}c-f.

We have also retrieved archival
Atacama Large Millimeter/submillimeter Array
(ALMA) observations of
CO\,($J=3-2$) and its associated 345\,GHz continuum from
program 2013.1.00740.S (PI: T.~Saito).
The images of the continuum-subtracted CO\,($3-2$)
emission (moment 0) and 345\,GHz continuum (extracted from line-free
channels) are shown in Fig.~\ref{maps}a-b \citep[see also][]{sai15}.
Both maps clearly delineate the elongated filament
and peak at the nuclear region of VV 114 E.
The IR images of this nucleus,
enlarged in Fig.~\ref{maps}c-f, show three main clumps,
which will be denoted as NE, SW-s1, and SW-s2 (panel e). These
correspond to sources ``a'', ``c'', and ``d'' in \cite{ric23},
and have 33\,GHz counterparts ``n6'', ``n4'', and ``n3'' as denoted
in \cite{son22}, respectively. The strongest source at 2.0, 5.6, and
7.7\,$\mu$m, the latter dominated by PAH emission, is SW-s1, but
NE dominates at longer wavelengths (Fig.~\ref{maps}f).
It is also worth noting that
SW-s1 is devoid of CO\,($3-2$) and 345\,GHz continuum, which surround
the clump (Fig.~\ref{maps}d,f) and peak at NE and SW-s2,
which also display the strongest emission at 33\,GHz
\citep{son22}.

We found a source of unusual CO and H$_2$O excitation in the SW-s2 core,
located just at the head of the 4\,kpc filament (magenta circle in
Fig.~\ref{maps}c-f). The {\it JWST} $2-10\,\mu$m spectrum extracted
from a $0.2''$ aperture at the position of the maximum band strength
is shown in Fig.~\ref{maps}g, where the spectral extent of the CO and
H$_2$O bands is indicated.

\subsection{The CO $v=1-0$ band}
\label{cobandobs}

The top panel of Fig.~\ref{bands} shows the $^{12}$CO (hereafter CO)
$v=1-0$ band of SW-s2 extracted from the NIRSpec high resolution
($\Delta v\sim90$\,km\,s$^{-1}$) grating G395H
in Fig.~\ref{maps}g. The baseline used
to subtract the continuum is shown and discussed in Appendix~\ref{basel}.
Absorption is observed in the P-branch up to $J=33$ (lower-level
energy of $E_{\mathrm{low}}=3000$\,K; the spectral features nearly
coincident with P(34) and P(35) are due to H$_2$O lines). Apparent
emission lines in the spectrum are due to H$_2$ $v=0-0$ S(8), S(9), and
S(10), and H I 7-5. The presence of two additional emission lines is
inferred from the relatively low absorption in the R(5) 
and R(24) lines, and are attributed to [\ion{K}{iii}]\,4.617\,$\mu$m and
[\ion{Mg}{iv}]\,4.488\,$\mu$m, respectively \citep{per23,gber23}.

As illustrated in the inserts of Fig.~\ref{bands}(upper),
the CO ro-vibrational lines are spectrally resolved and much broader
(full width at half maximum of
$\mathrm{FWHM}=200-270$\,km\,s$^{-1}$)
than and blueshifted relative to the rotational CO\,($3-2$)
profile ($\mathrm{FWHM}=105$\,km\,s$^{-1}$)
extracted from a similar aperture (radius $r=0.162''$).
Hereafter, we use the redshift of
the bulk of the gas in SW-s2, $z=0.02013$, derived from
the CO\,($3-2$) line fit.
Since adjacent R(J) lines are separated by $<520$\,km\,s$^{-1}$
(decreasing to $<400$\,km\,s$^{-1}$ for $J>22$), they
partially overlap forming a pseudo-continuum. Along the
P-branch, the CO line velocity separation increases 
(from 540 to 740\,km\,s$^{-1}$), but significant
absorption is still seen between adjacent P(6)-P(9), P(13)-P(15), and
P(20)-P(21) lines. At these wavelengths the $^{13}$CO and C$^{18}$O
ro-vibrational lines lie in between the $^{12}$CO lines, and
thus this absorption is attributed to the rare isotopologues.

A fuller view of the velocity profiles is given in
Fig.~\ref{covel}{\it left}, which shows
an energy level-velocity
diagram for the P-branch lines. Up to $J=16$
($E_{\mathrm{low}}=750$\,K) the peak absorption is blueshifted by
$70-100$\,km\,s$^{-1}$, but the blueshift increases to
$\approx170$\,km\,s$^{-1}$ for $J>20$. This is also illustrated
by the CO\,P(25) profile in Fig.~\ref{covel}{\it center}, which
shows little absorption at central velocities. The data thus
indicate the presence of two blueshifted components, with the most
excited component (denoted as the hot component, $H_C$) more
blueshifted than the less excited one (the warm component, $W_C$).
In addition, strong absorption is seen in the $J\leq3$
lines ($E_{\mathrm{low}}\le33$\,K) relative to $J=4$
(Fig.~\ref{bands}upper), indicating
the contribution to the absorption by a cold component ($C_C$),
presumably the quiescent gas that accounts for the CO\,($3-2$)
emission. The bulk of the absorption in the band
is due to the blueshifted $H_C$ and $W_C$, and therefore the lines
are all blueshifted relative to the transition labels in
Fig.~\ref{bands}(upper).

To compare the fluxes of the P(J) and R(J) lines arising from the
same $J$-level of the $v=0$ state, and due to line blending and
contamination by $^{13}$CO, we use instead the observed peak
absorption as shown in Fig.~\ref{peakabs}a,b. The P-R asymmetry
\citep[e.g.][]{gon02,per23,gber23}
is evaluated from the ratios of the peak absorption (in mJy) of these
pairs of lines, showing two different trends: for $J<10$,
$\mathrm{P(J)/R(J)}\lesssim1$, and for $J\ge10$, $\mathrm{P(J)/R(J)} >1$
increasing up to $\gtrsim1.5$ for the highest $J$.
The latter dependence is expected for
a background continuum that rises with $\lambda$,
as the P(J) (R(J)) transitions
have progressively longer (shorter) wavelengths.
A positive slope is
indeed observed in the continuum beween $4.4-5.0$\,$\mu$m 
(Fig.~\ref{maps}g), which is matched with a blackbody
radiation temperature of $T_{\mathrm{rad}}=390$\,K (Appendix~\ref{basel}).
The $\mathrm{P(J)/R(J)}\lesssim1$ values
found for relatively low $J$
require a mechanism that favors emission in the P-branch at the
expense of the R-branch, pointing toward line re-emission from the
flanks of the continuum source. CO absorption in the R-branch and emission
in the P-branch has been observed toward the disk of NGC~3256-S
\citep{per23} and previously toward
the galactic Orion BN/KL \citep{gon98}. No re-emision is however apparent
in the $H_C$.

Emission lines of other species in the spectrum of SW-s2 do not trace
the $H_C$. This is illustrated in Fig.~\ref{covel}{\it center}, which
compares the CO\,P(25) line profile with those of two H$_2$ lines, 
an H recombination line, and the [\ion{Fe}{ii}]5.3\,$\mu$m one.
All emission lines peak at central velocities,
with no blueshifted spectral feature similar 
to the CO\,P(25) line shape. The CO\,($3-2$) line neither
shows blueshifted emission, but a line wing at redshifted
velocities. A 3-components Gaussian fit
to the CO\,($3-2$) profile, fixing 
the central velocity of one component at $-160$\,km\,s$^{-1}$
(blue line in the CO\,($3-2$) panel of
  Fig.~\ref{covel}{\it center}),
establishes an upper limit for the flux of the $H_C$ in CO\,($3-2$)
of $<0.35$\,Jy\,km\,s$^{-1}$, equivalent to a luminosity of
$<7.4\times10^5$\,K\,km\,s$^{-1}$\,pc$^2$ and a gas mass of
$M_{\mathrm{gas}}(H_C)<4.4\times10^5\,M_{\odot}$
\citep[using $\alpha_{\mathrm{CO3-2}}\approx0.6\,M_{\odot}/(\mathrm{K\,km\,s^{-1}\,pc^2})$,][]{per23}.
The central velocity component has a CO\,($3-2$) flux of
$(17.4\pm0.1)$\,\,Jy\,km\,s$^{-1}$, which translates into 
$M_{\mathrm{gas}}(\mathrm{SW-s2})\sim3\times10^7\,M_{\odot}$
(using $\alpha_{\mathrm{CO3-2}}\approx0.8\,M_{\odot}/(\mathrm{K\,km\,s^{-1}\,pc^2})$)
and a beam-averaged column density of
$N_{\mathrm{H_2}}(\mathrm{SW-s2})\sim1\times10^{23}$\,cm$^{-2}$
($\Sigma_{\mathrm{gas}}(\mathrm{SW-s2})\sim2\times10^{3}$\,$M_{\odot}$\,pc$^{-2}$).

\subsection{The H$_2$O $\nu_2=1-0$ band}
\label{h2obandobs}

\begin{landscape}
\begin{figure}
\centering
\includegraphics[width=1.25\textwidth, angle=0]{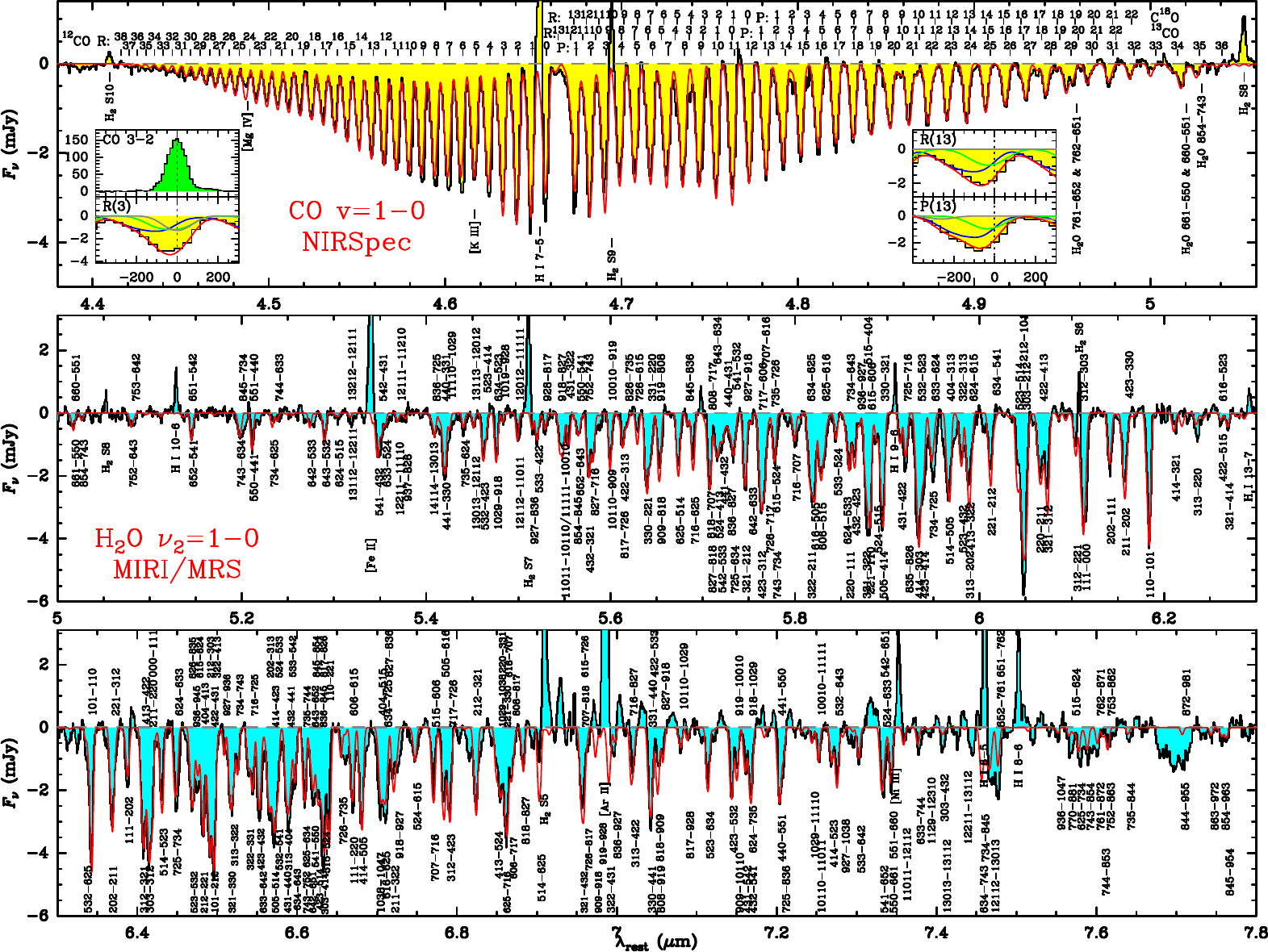}
\caption{CO $v=1-0$ (upper panel) and H$_2$O $\nu_2=1-0$ (middle and lower)
  bands in VV 114 SW-s2.
  Color-filled black histograms show the data, and the red lines show the total
  absorption predicted by our composite model. The inserts in the upper panel
  show the spectra
  of the CO 3-2 line observed with ALMA and several CO $v=1-0$ lines
  with the abscissae in units of velocity;
  the contributions of each of the three components (blue: hot;
  green: warm; gray: cold) are also shown. The labels for the H$_2$O
  $\nu_2=1-0\,J'_{K_a',K_c'}-J_{K_a,K_c}$ transitions, where the $'$ corresponds
  to the upper vibrational state, are indicated as
  $J'\,K_a'\,K_c'-J\,K_a\,K_c$ for readability. Note that all spectral
  features are blueshifted relative to the labels, because we use
  CO 3-2 as the velocity reference ($z=0.02013$).
}
\label{bands}
\end{figure}
\end{landscape}

The striking 5-8\,$\mu$m spectral region of SW-s2
  (Fig.~\ref{bands}b,c) contains $\sim150$ spectral features in absorption 
attributable to the H$_2$O $\nu_2=1-0$ band. Line identification, which
is based on the models described below (Sect.~\ref{anal}), indicates that
many of these feaures are 
produced by several blended
lines, and we estimate that $\sim280$ transitions of H$_2$O significantly
contribute to the observed spectrum. As shown in the energy level
diagram of Fig.~\ref{h2oenerlev}, 93 levels of the ground vibrational state
up to $13_{0,13}$ ($E_{\mathrm{low}}\approx2600$\,K) are involved. 

\begin{figure*}
   \centering
\includegraphics[width=17.0cm]{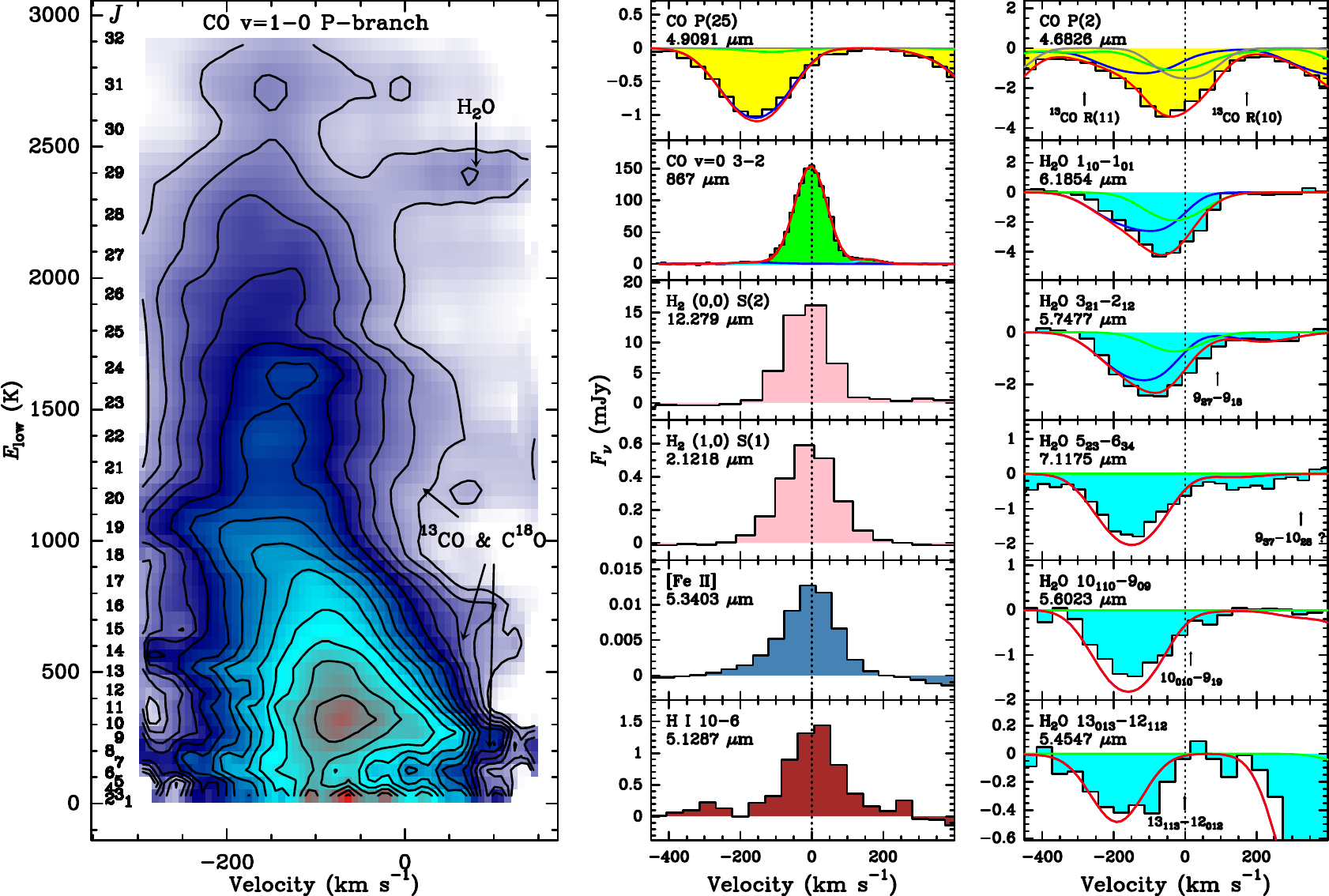}
\caption[]{{\it Left}: $E_{\mathrm{low}}$-velocity diagram for the CO P-branch
  lines in VV 114 SW-s2. Contamination by adjacent $^{13}$CO (together with
  C$^{18}$O) and H$_2$O lines is indicated on the red side of the profiles.
  The numbers on the left side indicate the rotational quantum number $J$ of
  the lower level.
  {\it Middle and right}: Comparison of the profiles of several
  CO and H$_2$O ro-vibrational lines (in yellow and light-blue, respectively) 
  and the profiles of CO $v=0\,J=3-2$ line observed with ALMA (green, with
  a Gaussian fit to the line profile in red),
  and of several lines of H$_2$ (pink) and hydrogen recombination line (brown).
  Prediction by our composite models for the ro-vibrational lines is
  overlaid.
}
\label{covel}%
\end{figure*}

The H$_2$O lines also show a blueshift relative to CO\,($3-2$), as illustrated
with five line profiles in Fig.~\ref{covel}{\it right}. As in the case of
CO, the low-excitation lines (such as the ground-state $\nu_2=1-0$
$1_{10}-1_{01}$ and the $3_{21}-2_{12}$ transitions)
are blueshifted by only $50-100$\,km\,s$^{-1}$ including
significant absorption at central velocities, while higher excitation lines
display blueshifts of $\gtrsim150$\,km\,s$^{-1}$. This is similar to
the 2 blueshifted velocity components, $W_C$ and $H_C$, that
also dominate the CO band.

To help characterize the complex H$_2$O band, we show in Fig.~\ref{h2obandelow}
the spectrum plotting $E_{\mathrm{low}}$ for the transitions that,
according to our models, significantly contribute to the forest
of absorption features. Most of the strongest spectral features
($<-2$\,mJy, in red) are dominated by low-excitation lines
($E_{\mathrm{low}}\lesssim500$\,K), but some high-excitation transitions
($E_{\mathrm{low}}\gtrsim800$\,K) also generate strong absorption
(see details in Appendix~\ref{h2olines}).

\section{Analysis}
\label{anal}

\subsection{The obscured continuum and the foreground extinction}
\label{contin}

We constrain the properties of the continuum
source behind the observed absorbing gas,
specifically its apparent temperature (i.e. once the
continuum has been extinguished by
intervening dust, $T_{\mathrm{app}}$),
the mid-IR extinction (which we characterize by the optical depth at 
$6\,\mu$m, $\tau_{\mathrm{6\mu m}}^{\mathrm{ext}}$), and the
intrinsic shape described by an equivalent blackbody temperature,
$T_{\mathrm{bck}}$.

Assuming  no reemission in
the CO band, the P(J)-to-R(J) peak absorption ratio is
$f_J\equiv\mathrm{P(J)/R(J)}=F_{\mathrm{c}}^{\mathrm{P(J)}}/
F_{\mathrm{c}}^{\mathrm{R(J)}}$ when the lines are optically thick, and
$f_J=(F_{\mathrm{c}}^{\mathrm{P(J)}}B_{\mathrm{lu}}^{\mathrm{P(J)}})/
(F_{\mathrm{c}}^{\mathrm{R(J)}}B_{\mathrm{lu}}^{\mathrm{R(J)}})$
in the optically thin limit. Here $F_{\mathrm{c}}^{\mathrm{P(J)}}$
($F_{\mathrm{c}}^{\mathrm{R(J)}}$) is the flux density of the 
continuum behind the absorbing gas
at the wavelength of the P(J) (R(J)) transition after foreground
extinction, and $B_{\mathrm{lu}}^{\mathrm{P(J)}}$ ($B_{\mathrm{lu}}^{\mathrm{R(J)}}$)
is the Einstein coefficient for absorption of radiation in the P(J) (R(J))
transition\footnote{We ignore here the slightly different spectral resolution
of NIRSpec at the wavelengths of the P(J) and R(J) lines, which is expected
to modify $f_J$ by less than 5\%.}.
Using blackbody emission at temperature $T_{\mathrm{app}}$ to
describe the $F_{\mathrm{c}}$ ratios,
we compare the theoretical curves
for $T_{\mathrm{app}}=330$, 400, and 450\,K, with the
observed $f_J$ values in Fig.~\ref{peakabs}b.
The latter are modulated by $^{13}$CO contamination
in the P-branch and severe blending in the R-branch, but the observed
positive slope favors the range
$T_{\mathrm{app}}=330-450$\,K for $J\ge10$.
This is consistent with the shape of the observed continuum
between $4.4-5.0$\,$\mu$m ($390$\,K, Appendix~\ref{basel}),
indicating that the continuum behind the
CO absorbing gas is,
once attenuated by foreground dust, similar in shape
to the observed continuum in Fig.~\ref{maps}g.

A similar approach cannot be reliably applied to the H$_2$O band due to
severe line blending, but an estimate of $T_{\mathrm{app}}$ can be obtained
by comparing the observed peak absorption of specific spectral features
across the band with the corresponding values obtained from models that
accurately account for line blending (Section~\ref{fit}). Up to 116
spectral features are considered in Fig.~\ref{peakabs}c-d
(and marked on the spectrum in Fig.~\ref{h2obandelow}),
covering most of the band extent.
LTE model results (Sect.~\ref{fit}) with $T_{\mathrm{app}}=400$ and 600\,K
are compared with data in Fig.~\ref{peakabs}c-d, indicating that
the $T_{\mathrm{app}}\sim400$\,K found for CO fails at the
long wavelength end of the H$_2$O band, because the H$_2$O P-branch lines 
at $>6.9$\,$\mu$m are systematically overestimated.
These features are better reproduced with $T_{\mathrm{app}}=600$\,K,
although the H$_2$O R-branch features at $<5.4$\,$\mu$m are then
overestimated. LTE models for the H$_2$O band favour
$T_{\mathrm{app}}\gtrsim500$\,K along the P-branch
($\lambda>6.8$\,$\mu$m).

The continuum behind the H$_2$O band thus flattens relative to that
behind the CO band. 
To explain this effect, invoking a distribution 
of dust temperatures ($T_{\mathrm{dust}}$) is disfavoured because
usually, the longer the wavelength, the lower the
$T_{\mathrm{dust}}$ being traced, but the effect found here is the opposite.
The increasing $T_{\mathrm{app}}$ with increasing $\lambda$ is
better ascribed to differential extinction.
The mid-IR extinction laws derived by \cite{ind05} and
\cite{chi06}, which we use hereafter
\citep[see also][]{xue16}, show a drop of $A_{\lambda}$
from near- to mid-IR wavelengths, but the drop is 
only $\approx8$\% along the CO band between $\sim4.4$ and 5\,$\mu$m.  
High values of $\tau_{\mathrm{6\mu m}}^{\mathrm{ext}}=2.5-3$ 
are then required to shape an intrinsic continuum with
$T_{\mathrm{bck}}\gtrsim550$\,K to $T_{\mathrm{app}}\sim400$\,K
(Appendix~\ref{extinc}).

The mid-IR spectral energy distribution (SED) in Fig.~\ref{maps}g
is consistent with the proposed extinction of the continuum associated
with the bands. We fitted the SED using a modified version of
the routine by \cite{don23} \citep[see also][]{gber22},
with a minimum number of three blackbody components, the PAHs, and
ice absorption. 
The fit in Fig.~\ref{maps}g fixes $T_{\mathrm{bck}}=550$\,K and
$\tau_{\mathrm{6\mu m}}^{\mathrm{ext}}=2.5$ for the component that
dominates the hot dust emission between 3.5 and
8\,$\mu$m (in light-blue). 
Between 4.4 and 5.0\,$\mu$m, $\tau^{\mathrm{ext}}$ varies between
2.7 and 2.9, shaping the continuum to an 
apparent $T_{\mathrm{app}}\approx420$\,K across
the CO band in rough agreement with the value derived from the
CO P-R asymmetry (see also Appendix~\ref{extinc}).
We adopt below these fiducial values for the models of the CO and
H$_2$O bands, noting however that
the source of mid-IR continuum must not necessarily be a blackbody,
but could be diffuse emission with an equivalent temperature as quoted above.
We will return to this point in Section~\ref{hc}.

\begin{table*}
  \caption{\label{tab}Properties of the components used to model
    the CO and H$_2$O fundamental bands in VV 114 SW-s2}
\centering
\begin{tabular}{ccccccccccc}
\hline\hline
 & $N_{\mathrm{CO}}$ & $N_{\mathrm{H_2O}}$ & $A_{\mathrm{CO}}$ & $A_{\mathrm{H_2O}}$ &
$V_{\mathrm{tur}}$ & $V_{\mathrm{gas}}$\tablefootmark{a} &
$d/R_{\mathrm{IR}}$\tablefootmark{b} &
$\Delta R/R_{\mathrm{IR}}$\tablefootmark{c} & $n_{\mathrm{H_2}}$  &
$T_{\mathrm{gas}}$ \\
& ($10^{19}$\,cm$^{-2}$) & ($10^{19}$\,cm$^{-2}$) & (pc$^{2}$) & (pc$^{2}$) &
(km\,s$^{-1}$) &  (km\,s$^{-1}$) &  &  & ($10^5$\,cm$^{-3}$) & (K) \\
\hline
$H_C$ & $1.8-3.5$ & $1.5-3.0$ & $0.14-0.25$ & $0.11-0.20$ &
60 & $300-50$ & 0 & $\lesssim0.5$ & \tablefootmark{d} & \tablefootmark{d} \\
$W_C$ & $0.5-1.2$ & $0.1-0.4$ & $0.11-0.20$ & $0.07-0.12$ &
60 & $30$ & $\gtrsim2$ & $-$ & $1-5$ & $300$  \\
$C_C$ & $\sim1$ & $-$ & $>0.14$ & $-$ &
60 & $0$ & $-$ & $-$ & $\sim1$ & $10$  \\
\hline
\hline
\end{tabular}
\tablefoot{
\tablefoottext{a}{Gas velocity varies across the shell in the $H_C$.}
\tablefoottext{b}{$d$ is the distance between the surface of the IR
  source and the inner part of the absorbing shell.}
\tablefoottext{c}{Thickness of the absorbing shell relative to the radius
  of the IR source.}
\tablefoottext{d}{Not constrained, because the $v=0$ levels are
  radiatively pumped.}
}
\end{table*}

\subsection{The best fit to the CO and H$_2$O bands}
\label{fit}

Our model for the bands includes the three components outlined
in Section~\ref{cobandobs}: the $H_C$ generating absorption in
the highest energy lines, the $W_C$ dominating the absorption at
lower energy, and the $C_C$, only for CO, which produces additional
absorption in the lowest $J$ lines. We use the code 
described in \cite{gon98}, which assumes spherical symmetry.
It has been updated to
generate modeled spectra convolved with the {\it JWST}/NIRSpec and
MIRI/MRS spectral resolution. The calculations include a careful
treatment of overlaps among lines of the same or different species.

We start modeling the $H_C$ by assuming thermalization of all
levels in the ground vibrational state at $T _{\mathrm{rot}}$ with no
significant population in the upper vibrational state.
The only free parameters in LTE models are $T _{\mathrm{rot}}$,
$N_{\mathrm{CO}}$ (including $^{13}$CO with $\mathrm{[^{12}CO]/[^{13}CO]=30-60}$),
$N_{\mathrm{H_2O}}$, and the gas velocity field.
We considered two different approaches for the latter:
$V_{\mathrm{field}}^{\mathrm{A}}$ denotes a
combination of turbulent velocity ($V_{\mathrm{tur}}=60$\,km\,s$^{-1}$)
and a velocity gradient across the shell, and $V_{\mathrm{field}}^{\mathrm{B}}$
uses a constant outflowing velocity of 190\,km\,s$^{-1}$ with a small
$V_{\mathrm{tur}}=20$\,km\,s$^{-1}$. In
the latter case, to match the observed linewidths, the profiles were
convolved with a Gaussian distribution of velocities, simulating an
ensemble of clumps with a velocity distribution along the line of sight.

Radiative transfer calculations were also performed to infer which physical
conditions are required to explain the observed excitation and absorption
in the $H_C$. Models with collisional excitation alone,
simulating shocked gas detached from the mid-IR source, 
and models including radiative pumping simulating gas in close
proximity to the mid-IR source behind it, were generated.
In the calculation of line fluxes,
the shell is truncated  on the flanks around the central source
as required to avoid line reemission \citep[see Fig.~8 in][]{gon14b}.
Statistical equilibrium calculations were performed using the 
rates by \cite{yan10} for collisional excitation of CO with H$_2$,
and those by \cite{dan11} for H$_2$O with H$_2$, both 
extrapolated to higher energy levels with the use of an
artificial neural network \citep{neu10}.

The $W_C$ is modeled as a shell of gas surrounding the mid-IR source
expanding at a velocity of 30\,km\,s$^{-1}$. A set of models was
generated by changing the distance from the surface of the central
source to the shell ($d$), the H$_2$ density and gas temperature
($n_{\mathrm{H_2}}$ and $T _{\mathrm{gas}}$), and the CO and H$_2$O column
densities ($N_{\mathrm{CO}}$, $N_{\mathrm{H_2O}}$). To match the line
profiles, $V_{\mathrm{tur}}=90$\,km\,s$^{-1}$ was used (giving
$\mathrm{FWHM}=150$\,\,km\,s$^{-1}$ for optically thin lines).

All combinations among the $W_C$ and $H_C$ model components were
cross-matched with the data by computing $\chi^2$ for the peak
absorption features (Fig.~\ref{peakabs}).
Minimization of $\chi^2$ gives the area that CO and H$_2$O
have in the plane of sky,
$A_{\mathrm{CO}}$ and $A_{\mathrm{H_2O}}$, in both components, and thus
the minimum area of the mid-IR continuum source behind it.
The derived parameters are listed in Table~\ref{tab}. Figures~\ref{bands},
\ref{covel}, and \ref{coprof} overlay the resulting best-fit model profiles
on the observed spectra, and Fig.~\ref{peakabs} compares the observed
peak absorption with the best-fit model predictions.
The models also give the strength of the continuum behind
each component, shown with dashed lines in Fig.~\ref{maps}g,
as required to match the absolute absorption fluxes.

\subsubsection{The Hot component}
\label{hc}

Extremely high column densities for both CO and H$_2$O are required to
explain the observed absorption strengths in the outflowing $H_C$.
The minimum values of both $N_{\mathrm{CO}}$ and $N_{\mathrm{H_2O}}$ are found in LTE
models with $V_{\mathrm{field}}^{\mathrm{B}}$:
$\approx1.5\times10^{19}$\,cm$^{-2}$ at $T_{\mathrm{rot}}\approx450-500$\,K. 
In CO, these are required to account for $(i)$ the small contrast between
the absorption strengths of intermediate $J=15-22$ and high $J>25$ lines,
indicating that the former lines are optically thick; $(ii)$ the absorption
in the $^{13}$CO P(18)-P(19) lines, further indicating optically thick
absorption in the same lines of $^{12}$CO, and $(iii)$ the blueshifted
wings observed in the low-$J$ lines.

In models with pure collisional excitation, additionally extremely high
densities of $>10^9$\,cm$^{-3}$ are needed to explain the H$_2$O excitation,
due to the high $A-$Einstein coefficients and critical densities
of the H$_2$O rotational transitions.
The combination of such high $n_{\mathrm{H_2}}$, $T_{\mathrm{gas}}$, and
$N_{\mathrm{CO,H_2O}}$ raises strong interpretation problems in the framework
of shock models where molecules are collisionally excited.
We have used the Paris-Durham shock code
\citep[e.g.][]{god19} to generate both $C-$type and $J-$type shock models
for a variety of pre-shock densities, magnetic fields, and shock velocities,
but failed to reproduce the required column densities of warm molecular gas.
Multiple ($>20$) shocked spots along the line of sight would be required
to attain the required columns at high $T_{\mathrm{gas}}$, but the
necessary pre-shock densities of $\gtrsim10^8$\,cm$^{-3}$ would in any case
indicate that the shock is produced very close to the source of IR
radiation.


On the other hand, the models that include
excitation by the mid-IR source at $\sim550$\,K
naturally yield $T_{\mathrm{rot}}\approx450-500$\,K as required
by LTE models, provided that the molecular
shell is in close proximity to the mid-IR source, and that the latter
is a blackbody in the mid-IR.
We conclude that the observed bands are closely associated with the
mid-IR continuum behind it and that the excitation of the CO and H$_2$O
rotational levels of the ground vibrational state is 
influenced by radiative excitation.
Indeed we find that the effect of mid-IR pumping is so
strong that results for the bands are basically insensitive to
$n_{\mathrm{H_2}}$ and $T_{\mathrm{gas}}$, and only depend on
$T_{\mathrm{bck}}$, $N_{\mathrm{CO,H_2O}}$, the width of the absorbing
shell, and the velocity field. In our best-fit model shown in
Figs.~\ref{bands}, \ref{covel}, \ref{peakabs}, and \ref{coprof},
$N_{\mathrm{CO}}=3.5\times10^{19}$\,cm$^{-2}$,
$N_{\mathrm{H_2O}}=3.0\times10^{19}$\,cm$^{-2}$,
$\Delta R/R_{\mathrm{IR}}=0.3$,
and $V_{\mathrm{field}}^{\mathrm{A}}$ is used.
As indicated in Table~\ref{tab}, the effective size of the 
absorbing shell is sub-pc.

The strength of the continuum behind the different components
as well as $A_{\mathrm{CO,H_2O}}$, depend
on the gas velocity field and shell width, and cannot be
constrained to better than 50\%.
Nevertheless, our best-fit models indicate that 
the sum of the continuum behind $H_C$ and $W_C$
as derived from CO is similar to the continuum due to
hot dust predicted
by our fit to the mid-IR emission (Fig.~\ref{maps}g).

\subsubsection{The Warm component}
\label{wc}

The $W_C$ most likely involves a mix of several components peaking
close to the reference velocity, such as material at the surface of
the IR source that is not covered by the $H_C$, gas in front of the
IR source more distant than the $H_C$, and gas on the flanks
that is illuminated and reemits preferently in the P($J<10$) lines
of CO. An additional difficulty comes from the fact that the
$H_C$ and $W_C$ partially overlap in velocity, and their relative
contribution to the moderate excitation lines is uncertain to
some extent. Here we explore whether gas in front of the IR source
can account for the remaining absorption unmatched by the $H_C$.
Because of the lower $W_C$ excitation relative to
the $H_C$, the warm gas is in this scenario at a
distance of $\gtrsim2R_{\mathrm{IR}}$ from the surface of the IR source.
The excitation is
in this case sensitive to $n_{\mathrm{H_2}}$ and $T_{\mathrm{gas}}$,
but is still also affected by the radiation field. Our results
enable the interpretation of the $W_C$ as an extension of the
$H_C$ further from the IR source, and could represent the shock
produced by the inner outflow ($H_C$) as it sweeps out the
surrounding ambient gas ($C_C$), or the remnant of a previous 
outflowing pulse. The latter is favored due to
the discontinuity in excitation between the $H_C$ and the
$W_C$, as readily seen in the overall CO band shape.

\subsubsection{The Cold component}
\label{cc}

As noted in Section~\ref{cobandobs}, the CO band shows strong
absorption in the $J\leq3$ lines, indicating
the presence a cold component ($C_C$) in front of the continuum
source. $C_C$ is modeled with a screen approach as a cold slab 
with $N_{\mathrm{CO}}=1\times10^{19}$\,cm$^{-2}$,
$n_{\mathrm{H_2}}=1\times10^{5}$\,cm$^{-3}$, and $T_{\mathrm{gas}}=10$\,K.
These parameters are relatively uncertain because $C_C$ is most
likely absorbing a continuum that has already been partially
absorbed by the $W_C$ at similar velocities, and the model
does not simulate this effect. $C_C$ also shows up in the
$^{13}$CO P(1) and P(2) lines, blueshifted relative to 
$^{12}$CO P(13) and P(14) (Fig.~\ref{coprof}).

\begin{figure*}
   \centering
\includegraphics[width=17.0cm]{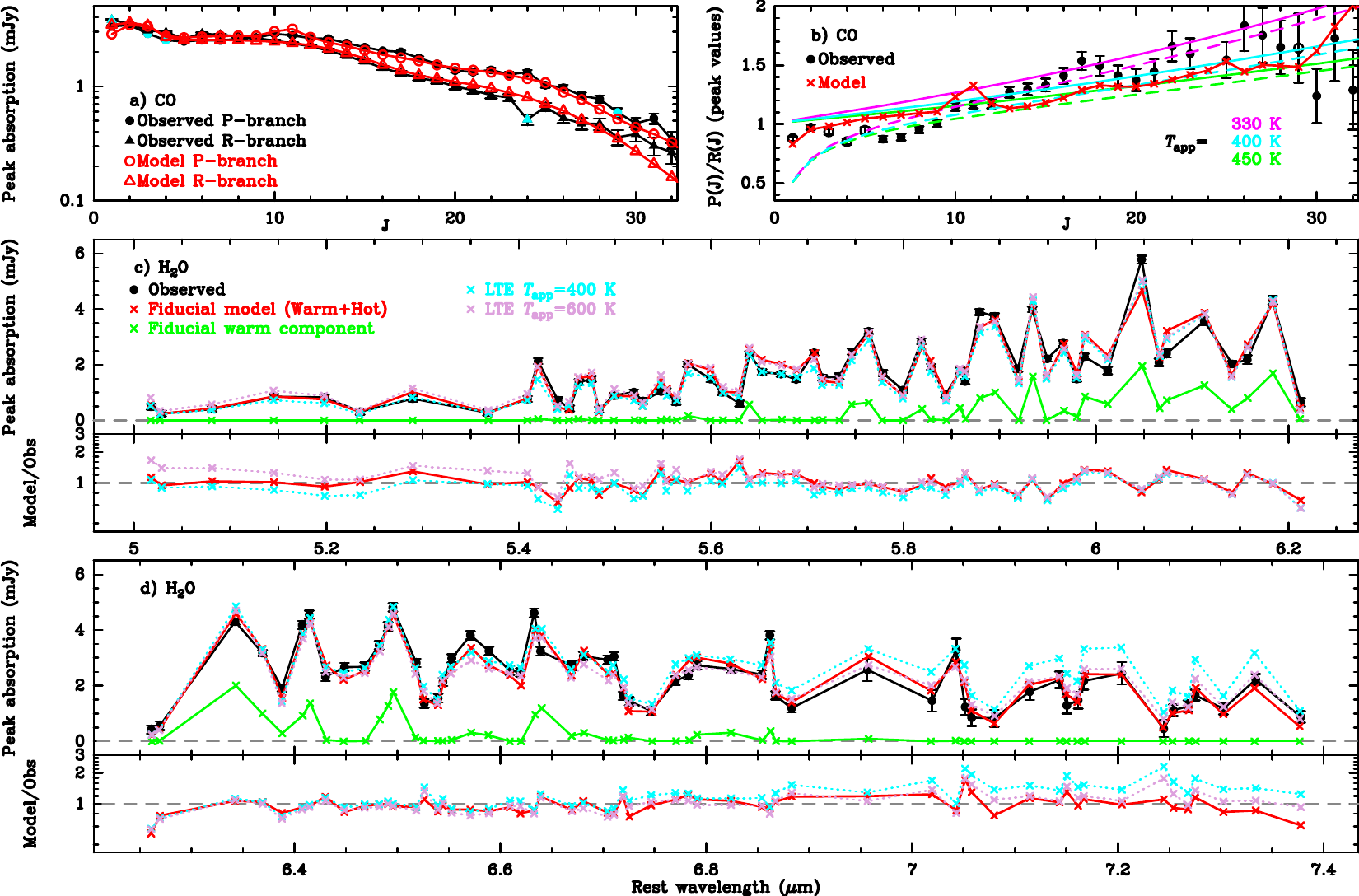}
\caption[]{{\bf a)} Peak absorption values of the CO
  P(J) (circles) and R(J)
  (triangles) lines. Black filled symbols indicate data, and red symbols show
  the model prediction. Light-blue markers indicate contaminated lines.
  {\bf b)} CO P-R asymmetry for the peak absorption values.
  Black circles indicate data, with opened symbols indicating doubtful ratios
  due to contaminating lines other than $^{13}$CO. Red crosses show
  model results. The magenta, light-blue, and green curves display
  the expected trends for several $T_{\mathrm{app}}$ values in the
  optically thick (solid) and optically thin (dashed) limits.
  {\bf c-d)}  Comparison between the H$_2$O peak absorption of 
  116 spectral features (black circles) and our fiducial model.
  The contribution by the warm
  component is shown in green, and the total predicted absorption (warm+hot
  components) is shown in red. LTE model results with
  $T_{\mathrm{app}}=400$ and 600\,K, $T_{\mathrm{rot}}=450$\,K, and
  $N_{\mathrm{H_2O}}=3\times10^{19}$\,cm$^{-2}$ (the same $N_{\mathrm{H_2O}}$
  as in the fiducial model)
  are also compared with data. The model-to-observed peak absorption 
  ratios are also displayed.
  Errorbars in this figure do not include uncertainties
    from continuum subtraction.
}
\label{peakabs}%
\end{figure*}

\section{Discussion and conclusions}
\label{disc}


\subsection{Energetics}
\label{ener}

The extremely high column densities found for the $H_C$ indicate that
the observed absorption is most likely dust-limited. At 6\,$\mu$m the
mass absorption coefficient is $\approx15$ times higher than at 
100\,$\mu$m. Using the $N_{\mathrm{H}}-\tau_{\mathrm{100\mu m}}$ relationship
in \cite{gon14}, we expect $\tau_{\mathrm{6\mu m}}=1$ for
$N_{\mathrm{H}}\sim10^{23}$\,cm$^{-2}$, and our derived CO and H$_2$O
column densities in the $H_C$ of $\mathrm{a\,few}\,\times10^{19}$\,cm$^{-2}$
roughly give the quoted $N_{\mathrm{H}}$ for abundances of
$\mathrm{a\,few}\,\times10^{-4}$. We could thus be still missing
outflowing material behind the curtain of dust.

Adopting $N_{\mathrm{H}}\sim10^{23}$\,cm$^{-2}$, the gas mass
of the outflowing $H_C$ is
$M_{H_C}=\mu m_{\mathrm{H}} A_{\mathrm{CO}}N_{\mathrm{H}}\approx170\,M_{\odot}$,
where $\mu=1.36$ accounts for elements other than hydrogen and we
have used the minimum value of $A_{\mathrm{CO}}$ in Table~\ref{tab}.
In spite of the high column densities, the small size 
yields a low mass that makes the $H_C$ undetectable in the CO
rotational lines at millimeter wavelengths (Sect.~\ref{cobandobs}).

The mass outflow, momentum, and energy rates can be estimated as
time-averaged or instantaneous values, depending on whether
the outflowing shell radius $R$ or the shell width $\Delta R$ 
are used in the equations \citep{rup05,gon17,vei17}. Adopting here
the (higher) instantaneous values with $\Delta R\sim0.1$\,pc,
we obtain $\dot{M}_{H_C}\sim0.3\,M_{\odot}\,\mathrm{yr^{-1}}$,
$\dot{P}_{H_C}\sim3.6\times10^{32}$\,dyn, and
$\dot{E}_{H_C}\sim3.2\times10^{39}$\,erg\,s$^{-1}$. These estimates
are likely lower limits, 
but indicate rather moderate momentum and energy rates as compared
with the radiation pressure and IR luminosity (see below),
$\dot{P}_{H_C}/(L_{\mathrm{IR}}/c)\sim0.4$ and
$\dot{E}_{H_C}/L_{\mathrm{IR}}\sim10^{-4}$.
Therefore, radiation pressure can drive the observed outflow.

\subsection{The origin of the bands}
\label{origin}

A crucial point of our analysis is that the rotational levels of the
ground vibrational state of CO and H$_2$O are radiatively excited
by the mid-IR continuum, and thus this
continuum is optically thick -i.e. a true mid-IR blackbody with
temperature $T_{\mathrm{bck}}\approx550$\,K. On the other hand,
the extreme CO and H$_2$O excitation and specific
  kinematics of the $H_C$ indicate a common environment,
and thus the area covered by the outflow in the plane of sky
is most likely of the same order as the effective areas listed in
Table~\ref{tab}. This implies that the obscured hot blackbody
emission forms a coherent, connected structure.

The size of this coherent bright object
  can be estimated from the sum of the areas
  $A_{\mathrm{CO}}(H_C)+A_{\mathrm{CO}}(W_C)$, giving
  $R_{\mathrm{IR}}=0.31$\,pc, which assumes that both areas
  do not overlap.
  We estimate the luminosity of this sub-pc structure as
  $L_{\mathrm{IR}}=(2-4)\times(A_{\mathrm{CO}}(H_C)+A_{\mathrm{CO}}(W_C))
  \sigma T_{\mathrm{bck}}^4$,
  where the numerical factors correspond to a disk seen face-on
  and a sphere, respectively, giving
  $\sim(0.8-1.7)\times10^{10}\,L_{\odot}$. 
  The luminosity surface density is given by
  $\Sigma_{\mathrm{IR}}=\sigma T_{\mathrm{bck}}^4
  \approx1.3\times10^{10}$\,$L_{\odot}$\,pc$^{-2}$.
  The areas and $L_{\mathrm{IR}}$ scale with foreground extinction
  as $\propto\exp\{\tau_{\mathrm{6\mu m}}^{\mathrm{ext}}-2.5\}$.
  Because of possible departures from blackbody emission, lower
  $T_{\mathrm{dust}}$ behind the area covered by the $W_C$, and beaming
  effects (see below), we write
  \begin{equation}
    L_{\mathrm{IR}}\approx(1\pm0.5)\times10^{10}\,
    \exp\{\tau_{\mathrm{6\mu m}}^{\mathrm{ext}}-2.5\}\,L_{\odot}.
  \end{equation}

We remark that the estimated value 
$\Sigma_{\mathrm{IR}}\sim10^{10}$\,$L_{\odot}$\,pc$^{-2}$, which is compared
below with the values in other sources, is not based
on direct measurements of the source size. The SW-s2 clump, as
seen with the ALMA angular resolution of $0.14''\approx60$\,pc, cannot be
used to constrain the properties of the extremely compact source of
mid-IR emission; indeed, the measured 345\,GHz continuum flux density
(Fig~\ref{maps}b,f) is
$\approx1$\,mJy\,beam$^{-1}$, and a blackbody of 550\,K and radius
$0.31$\,pc gives $0.08$\,mJy.
The CO and H$_2$O molecular bands are used as a surrogate for spatial
resolution, enabling us to identify high
concentrations of hot dust and to estimate $\Sigma_{\mathrm{IR}}$ 
on the assumption of a connected structure for the associated
mid-IR emission.

We compare the physical properties derived
for VV~114 SW-s2 with those found for the galactic massive
($\sim45\,M_{\odot}$), luminous ($1\times10^5\,L_{\odot}$) isolated
protostar AFGL 2136 IRS~1, which has been also observed 
in the CO $v=1-0$ band \citep{mit90} and
in the H$_2$O $\nu_2$, $\nu_1$ and
$\nu_3$ fundamental bands from the ground
and with the EXES instrument on SOFIA, with all
IR lines in absorption \citep{ind13,ind20,bar22}.
The bands and the associated near- to
mid-IR continuum probe a Keplerian disk around the
protostar. 
A large-scale bipolar outflow is observed in CO at millimeter
wavelengths with velocities $\lesssim20$\,km\,s$^{-1}$
\citep{kas94}, but not in the ro-vibrational lines of CO or H$_2$O.
The inner disk has been resolved in the
H$_2$O\,$\nu_2=1-1\,5_{50}-6_{43}$ line with ALMA,
giving a radius of 120 au \citep{mau19}.
Using $R_{\mathrm{IR}}=0.31$\,pc and
$L_{\mathrm{IR}}=1.7\times10^{10}\,L_{\odot}$ for
VV~114 SW-s2, we find that the ratio of
SW-s2 to AFGL 2136 luminosities,
$\sim1.7\times10^5\exp\{\tau_{\mathrm{6\mu m}}^{\mathrm{ext}}-2.5\}$,
is similar to the ratio of squared sizes,
$2.8\times10^5\exp\{\tau_{\mathrm{6\mu m}}^{\mathrm{ext}}-2.5\}$,
indicating similar continuum brightness. Indeed, the 
rotational temperature derived from the H$_2$O absorption in
AFGL 2136 IRS~1, $\sim520$\,K 
\citep{ind20}, is remarkably close to our derived
$T_{\mathrm{bck}}\approx550$\,K. Therefore, VV~114 SW-s2 behaves
in continuum brightness as the inner hot disk around a
high-mass protostar
\citep[another aspect of the analogies found between protostars
and buried galaxy nuclei;][]{dud97,gor23}, but with a
projected area
$\gtrsim10^5\exp\{\tau_{\mathrm{6\mu m}}^{\mathrm{ext}}-2.5\}$
times larger, and with the CO and
H$_2$O fundamental bands, rather than tracing a stationary
disk/torus, outflowing at
$V_{\mathrm{out}}\approx180$\,km\,s$^{-1}$.
While the flow timescale is only
$R_{\mathrm{IR}}/V_{\mathrm{out}}\sim1.7\times10^3$\,yr,
an upper limit is constrained by
$M_{\mathrm{gas}}(H_C)/\dot{M}_{H_C}\lesssim10^6$\,yr.
Whatever is the source of power heating the dust and driving
the outflow in VV~114 SW-s2, it is caught in a very
early phase of evolution where feedback has not yet been able to
clear the natal cocoon, at least on the side
facing the observer. 

Another source of comparison, with a spatial scale more akin to
VV~114 SW-s2, is the proto-super star cluster 13
(p-SSC13) in NGC~253. \cite{rico22} observed this source with
high angular resolution ($0.02''\approx0.4$\,pc) in the millimeter
continuum and in a suite of HC$_3$N rotational lines from the
ground and excited vibrational states, up to $\nu_6=2$ ($>1500$\,K).
They found that p-SSC13, with radius $\approx1.5$\,pc,
is extremely buried ($\sim10^{25}$\,cm$^{-2}$), and trapping of
continuum photons raises the inner $T_{\mathrm{dust}}$ such
that the pumping of the HC$_3$N excited vibrational states
is very effective \citep[i.e. the greenhouse effect,][]{gon19}.
As a result, the inferred $L_{\mathrm{IR}}$ and $\Sigma_{\mathrm{IR}}$ 
remain moderate ($\approx10^8\,L_{\odot}$ and
$\approx10^7\,L_{\odot}$\,pc$^{-2}$, respectively).
We note that, while the millimeter high-excitation
HC$_3$N lines are not extinguished and thus probe the inner
thermal structure of p-SSC13, the CO and H$_2$O bands seen in
absorption and blueshifted in VV~114 SW-s2 are surface tracers,
revealing the actual
$\Sigma_{\mathrm{IR}}\sim1\times10^{10}\,L_{\odot}$\,pc$^{-2}$
associated with the outflowing gas.

\begin{figure}
   \centering
\includegraphics[width=8.0cm]{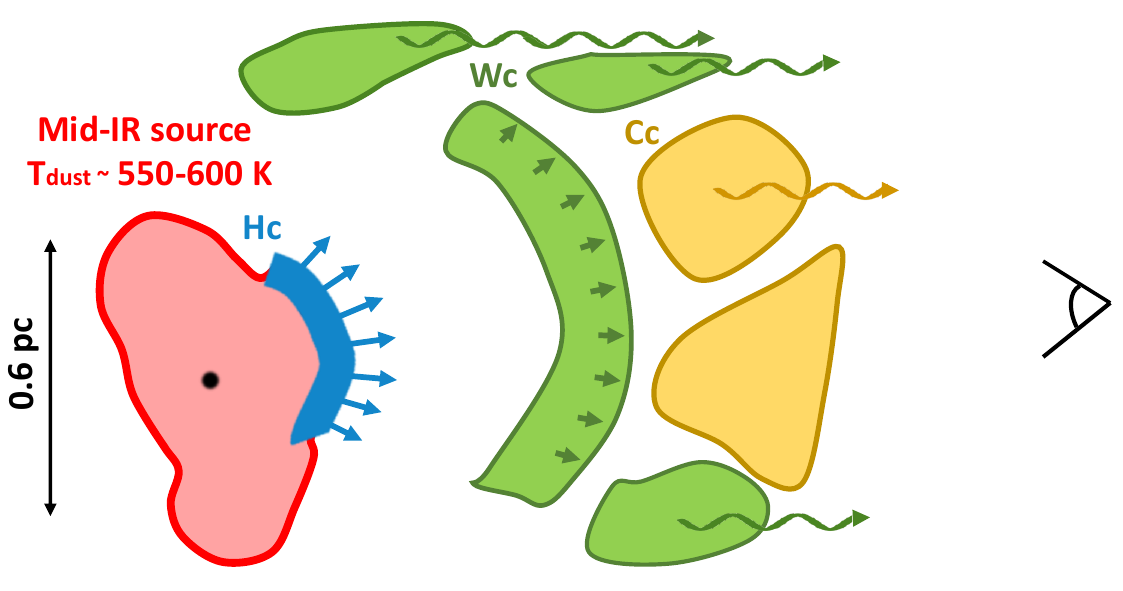}
\caption[]{Sketch of VV 114~SW-s2. The 3 components used in
  the modeling, $H_C$, $W_C$, and $C_C$, are represented in
  blue, green, and orange, respectively. The outflow takes places in
  the polar direction.
}
\label{sketch}%
\end{figure}

\cite{rico22} found that the
$\Sigma_{\mathrm{IR}}\sim10^7\,L_{\odot}$\,pc$^{-2}$ value derived for
p-SSC13 ocuppies an intermediate position between galactic
star-forming regions and bright (U)LIRGs
\citep[where values up to $\sim10^{8.3}\,L_{\odot}$\,pc$^{-2}$
have been derived for the nuclear $<100$\,pc regions,][]{per21}.
Since massive star formation in p-SSC13 
is taking place very efﬁciently, its $\Sigma_{\mathrm{IR}}$
was taken as a proxy for the most extreme starburst 
at $\gtrsim0.1$\,pc spatial scales. 
On this basis, the $\Sigma_{\mathrm{IR}}\sim10^{10}\,L_{\odot}$\,pc$^{-2}$
of VV~114 SW-s2 can be most easily understood as a 
black hole (BH) in a very early stage of accretion
and feedback.

Our identification of VV~114 SW-s2 as an AGN
  finds support in the observation of the CO $v=1-0$ band in
  IRAS~08572$+$3915~NW by \cite{oni21}. This source is a well known
  AGN-dominated ULIRG with $L_{\mathrm{AGN}}\sim10^{12-13}\,L_{\odot}$
  \citep{vei09,efs14}. The observations covered CO band lines from 
  R(26) to P(19), showing one very excited and broad velocity component
  blueshifted by $-160$\,km\,s$^{-1}$ relative to systemic, presumably
  arising from the innermost region of the torus surrounding the AGN
  \citep[component (a) in][]{oni21}. The blueshift of this component
  is very similar to that found for the $H_C$ of SW-s2. Concerning
  the excitation, the peak absorption observed in IRAS~08572$+$3915~NW
  decreases from $\approx30$\% of the observed continuum for the R(20)
  line to $\approx10$\% for the R(25) line, while in SW-s2 the absorption
  decreases more slowly with increasing $J$, from $\approx20$\% (R(20))
  to $\approx12$\% (R(25)) (Fig.~\ref{coprof}). Thus CO appears to be
  similarly or even more excited in SW-s2 than in IRAS~08572$+$3915~NW.

\subsection{Lifetime constraints and the mass of the BH}
\label{mbh}

Assuming Eddington luminosity,
the BH mass is $M_{\mathrm{BH}}\sim3\times10^5
\exp\{\tau_{\mathrm{6\mu m}}^{\mathrm{ext}}-2.5\}\,M_{\odot}$, but we can
refine this estimate because the derived luminosity is not
necessarily Eddington and, as shown below, constraining the BH evolving
time $t$ is equivalent to constraining $M_{\mathrm{BH}}$.
To apply a limit to the BH growing time in SW-s2
we assume that the BH has evolved locally from the seed
(with adopted initial mass $M_{\mathrm{BH,0}}=100\,M_{\odot}$)
so that the BH evolving time cannot be longer than
the lifetime of its s2 host.
The high concentration of CO cold gas at the s2 clump (Fig.~\ref{maps}a,d)
and the high column densities of hot molecular gas strongly suggest that this
is the case, supporting also the
assumption that the main channel of BH growth in SW-s2 is gas accretion.

A first approach for setting the lifetime of the s2 clump is from
the works by \cite{lin21,lin23}, who studied in a large number of clumps
of VV 114, including the overlap region, the colors from
optical/UV {\it Hubble} to near-IR {\it JWST} data. By comparing the observed
colors with predictions from evolutionary models of stellar population,
they found that the most embedded star clusters, undetected in the optical/UV,
have lifetimes $t\lesssim5$\,Myr.
While the SW-s2 clump undoubtedly belongs to the category of strongly embedded
sources, it may not represent the majority of star clusters
in the \cite{lin23} sample, because of its location at the head of the
filament (Fig.~\ref{maps}a).
The merger-induced shocked gas at the overlap region \citep{sai15} may lose
angular momentum and flow along the filament, replenishing the nuclear
region of VV 114 E and specifically the SW-s2 clump.
Unfortunately, the method used by \cite{lin23} to estimate the age of the
clusters, which mostly relies on the blue to red supergiant transition
traced by the $\mathrm{F150W-F200W}$ color, cannot be applied to SW-s2
because its F200W ($2\,\mu$m) emission is at the background level
(Fig.~\ref{maps}c).

On the other hand, it is clear from our analysis of the spectra and
timescales that the observed region is currently evolving very quickly,
and that the $H_C$ may indeed represent
a peak of activity. On longer time
scales the head of the VV 114 E appears to be evolving quickly as well, with
separated spots that have not yet coalesced. In addition, the s2 clump
appears off-center with brighter regions in NE and SW-s1, suggesting that
s2 is relatively young. The depletion timescale for the dense gas in s2
is $\sim6$\,Myr \citep[the S5 ``box'' in][]{sai15}.
Simulations of super star cluster formation and evolution \citep{ski15},
which is the most likely environment for BH seed formation and growth
\citep[e.g.,][]{por02,gur04,gos12}, indicate that the bulk of the initial
reservoir of gas is locked into stars or ejected after $\sim5t_{\mathrm{ff}}$,
($t_{\mathrm{ff}}$ is the initial free-fall time), and thus the gas surface
density significantly decreases afterwards. Using the extraction radius
of the CO (3-2) profile in SW-s2 ($r=70$\,pc) and the gas mass inferred 
from the line flux ($\sim3\times10^7\,M_{\odot}$),
$t_{\mathrm{ff}}=1.8\times(1+M_*/M_{\mathrm{gas}})^{-1/2}$\,Myr and we thus expect
$t\lesssim9$\,Myr in view of the still embedded stage of the clump.
All in all, we propose a fiducial conservative timescale for
the BH growth in SW-s2 of $t\sim10$\,Myr, with an uncertainty of a factor
$\sim2$ to accommodate more extreme cases \citep[see Fig~4 in][]{lin23}.

The simple BH growth model in Appendix~\ref{mbhderived} shows that the
quoted time constraint translates
into an $e-$folding time of $t_0\sim2^{+2}_{-1}$\,Myr, and using the relationship
between AGN luminosity and accretion rate by \cite{wat00}
\citep[also used by][]{toyi21}, we arrive at
\begin{equation}
  M_{\mathrm{BH}} \sim
    \frac{3.7\times10^4\,M_{\odot}}{1-0.244 \ln t_{0}} \,
  \exp\{\tau_{\mathrm{6\mu m}}^{\mathrm{ext}}-2.5\},
\end{equation}
  where $t_0$ is in Myr. Because of the weak dependence on $t_0$, this
  result favors an IMBH in SW-s2:
\begin{equation}
  M_{\mathrm{BH}}=(4.5\pm2.4)\times10^4
  \exp\{\tau_{\mathrm{6\mu m}}^{\mathrm{ext}}-2.5\}\,M_{\odot},
  \label{eq:mbhref}
\end{equation}
where the quoted error includes uncertainties in $t_0$ and $L_{\mathrm{IR}}$
added in quadrature.
Both the current accretion rate
($\dot{M}_{\mathrm{BH}}/\dot{M}_{\mathrm{Edd}}\sim25$) and luminosity
($L/L_{\mathrm{Edd}}\sim7$) are super-Eddington, with
$\dot{M}_{\mathrm{BH}}\sim2.4\times10^{-2}
\exp\{\tau_{\mathrm{6\mu m}}^{\mathrm{ext}}-2.5\}$\,$M_{\odot}$\,yr$^{-1}$.
If the $H_C$ represents a peak
of activity with the current $L/L_{\mathrm{Edd}}$ above the time-averaged value,
$M_{\mathrm{BH}}$ will be lower than in eq.~(\ref{eq:mbhref}).

Detailed 3D simulations of super-Eddington accretion onto IMBHs,
limited to low metallicities ($Z\leq0.1 Z_{\odot}$) and
$M_{\mathrm{BH}}=10^4\,M_{\odot}$, have been performed by \cite{toyi21}
\citep[see also][for extreme BH
growth in metal poor environments]{ina16}.
These models yield
super-Eddington accretion if the dusty disk becomes
optically thick to ionizing radiation, as is most likely the case for
SW-s2. \cite{toyi21} calculated, over an elapsed time 
of $\lesssim10$\,Myr,
both the time-averages of mass accretion rate onto the BH
($\langle \dot{M}_{\mathrm{BH}}\rangle$) and mass outflow rate
($\langle \dot{M}_{\mathrm{out}}\rangle$), as a function of the
Eddington-normalized
mass injection rate from the outer source boundary
($\dot{M}_{\mathrm{in}}/\dot{M}_{\mathrm{Edd}}$).
Their $Z=0.1\,Z_{\odot}$ model that better resembles our results for
SW-s2 has $\dot{M}_{\mathrm{in}}/\dot{M}_{\mathrm{Edd}}=10$, yielding
$\langle\dot{M}_{\mathrm{out}}\rangle / \langle\dot{M}_{\mathrm{BH}}\rangle\approx6$.
The outflow, driven by radiation pressure on dust grains
(as for the $H_C$, Section~\ref{ener}), takes places in
the polar direction and is time-variable
(as also favored for SW-s2, Section~\ref{wc}).
Our result for the mass outflow rate of the $H_C$
($\dot{M}_{H_C}\gtrsim0.3\,M_{\odot}\,\mathrm{yr^{-1}}$) yields
a comparable $\dot{M}_{H_C} / \dot{M}_{\mathrm{BH}}\gtrsim12$,
although models with higher metallicities \citep{sai15}
are required to refine this comparison.
According to \cite{shi23}, the global SW-s2 
  conditions required for a significant chance of 
  runaway BH growth via gas accretion are met
  ($\Sigma_{\mathrm{gas}}\gtrsim10^3$\,$M_{\odot}$\,pc$^{-2}$,
  $M_{\mathrm{gas}}>10^6$\,$M_{\odot}$).

VV 114 SW-s2 possibly exemplifies the 
IMBH to SMBH transition
in the evolved Universe, diagnosed in the mid-IR
as sub-pc fully-enshrouded cocoons of
hot dust where outflows have not yet been able to evacuate
the polar regions to form an open torus.
The cartoon in Fig.~\ref{sketch} shows a possible
schematic geometry of the relative locations
of the components observed in the CO and H$_2$O bands.
The {\it JWST} observations of VV 114 E
  suggest that IMBHs/SMBHs can be formed throughout
  the merging process in distinct nuclear
  clumps of the same merging galaxy, and could
  coalesce afterwards.

The unprecedented sensitivity and spectral resolution of
{\it JWST} enable detection of this type of buried object
in extragalactic sources. Ongoing and future observations
will reveal how common they are, and how they evolve.

\begin{acknowledgements}
  The authors acknowledge the DD-ERS teams for developing their observing
  program with a zero--exclusive--access period.
  EG-A acknowledges grants PID2019-105552RB-C4 and PID2022-137779OB-C41
  funded by the Spanish MCIN/AEI/10.13039/501100011033.
  IGB acknowledges support from STFC through grant ST/S000488/1 and
  ST/W000903/1.
  MPS acknowledges funding support from the Ram\'on y Cajal programme of
  the Spanish Ministerio de Ciencia e Innovaci\'on (RYC2021-033094-I).
  D.A.N. acknowledges funding support from USRA grant SOF08-0038.
  This work is based on observations made with the NASA/ESA/CSA James Webb Space
  Telescope. The data were obtained from the Mikulski Archive for Space
  Telescopes at the Space Telescope Science Institute, which is operated by
  the Association of Universities for Research in Astronomy, Inc., under
  NASA contract NAS 5-03127 for JWST; and from the European JWST archive (eJWST)
operated by the ESAC Science Data Centre (ESDC) of the European Space
Agency. These observations are associated with program \#1328.
This paper makes use of the ALMA data ADS/JAO.ALMA\#2013.1.00740.S.
ALMA is a partnership of ESO (representing its member states), NSF (USA) and
NINS (Japan), together with NRC (Canada) and NSC and ASIAA (Taiwan) and KASI
(Republic of Korea), in cooperation with the Republic of Chile. The Joint ALMA
Observatory is operated by ESO, AUI/NRAO and NAOJ. The National Radio
Astronomy Observatory is a facility of the National Science Foundation
operated under cooperative agreement by Associated Universities, Inc.
\end{acknowledgements}

\bibliographystyle{aa}
\bibliography{refs}

\begin{appendix}

\section{Data reduction}
\label{reduc}

  {\it JWST} imaging and integral field spectroscopy (IFU) of VV~114
  were taken  as part of the Director’s Discretionary Early
Release Science (DD-ERS) Program ID \#1328 (PI: L.~Armus and
A.~Evans). MIRI-MRS ($4.9-28.1$\,$\mu$m) and NIRSpec-IFU data
with the high spectral resolution ($R\sim2700$) gratings were
retrieved, as well as the fully reduced and calibrated NIRCam
F200W, and MIRI F560W, F770W, and F1500W images.

The NIRSpec-IFU data of the CO band were obtained with the
grating/transmission filter pair G395H/F290LP, using
the NRSIR2RAPID readout
with 18 groups and a small cycling dither pattern (4 points).
The spaxel size is $0.1''$. The observations were processed
using the JWST Calibration pipeline (version 11.16.20). All
details of the reduction procedure are given in \cite{gber23} and
\cite{per23}. 

The MIRI-MRS data of the four channels
were reduced using the 1.11.0 pipeline version and
calibration 1095 of the Calibration Reference Data System.
The pixel size is $0.196''$ for channels 1 and 2, and
increases to $0.245''$ for channel 3 with likely
contamination by the SWS-s1 core (Fig.~\ref{maps}).
For this reason, we show the spectrum in Fig.~\ref{maps}g
up to 10\,$\mu$m.
We primarily followed the standard MRS pipeline
procedure \citep[e.g.,][and references therein]{lab16} 
and the same steps as described in \citet{gber22} and
  \citet{per22} to reduce the data.
Some hot\slash cold pixels are not identified by the current
pipeline, so we added some extra steps as described in
\citet{per23} and \citet{gber23b} for NIRSpec and MRS,
  respectively.

\begin{figure*}
   \centering
\includegraphics[width=15cm]{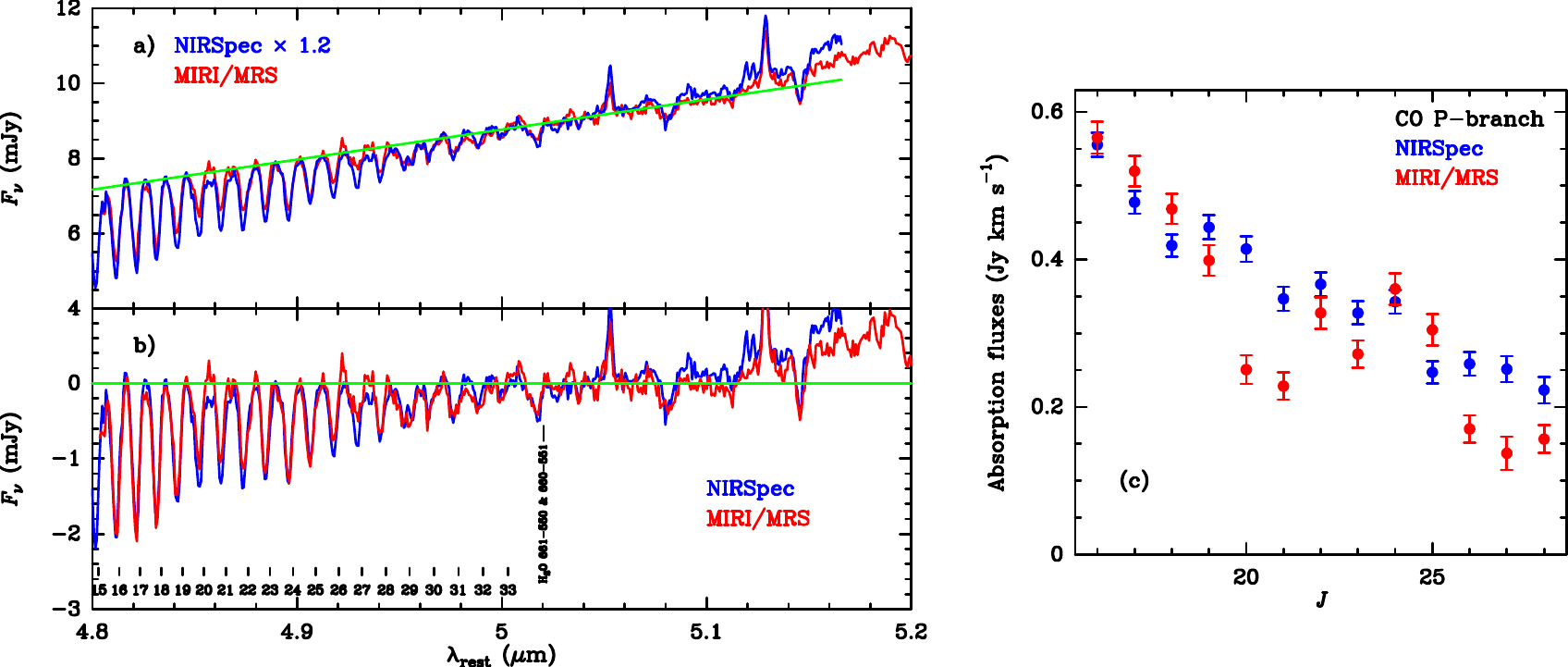}
\caption[]{{\it a)} NIRSpec G395H and MIRI/MRS CH1-Short
  spectra of VV114 SW-s2 in the $4.8-5.2$\,$\mu$m overlap region.
  The NIRSpec spectrum has been scaled up by a factor 1.2 to match
  the MIRI/MRS continuum. The green line is the fiducial baseline
  used for the CO band.
  {\it b)} Continuum-subtracted spectra, where the
  NIRSpec spectrum has not been scaled up.
  {\it c)} CO P(16)-P(28) line fluxes from Gaussian fits to
  the continuum-subtracted spectra of panel b).
}
\label{conirspecmiri}%
\end{figure*}

  In order to obtain a continuous near- to mid-IR spectrum of VV 114 SW-s2
(Fig.~\ref{maps}g), the flux densities of all NIRSpec bands were
  scaled up by a factor of 1.2, due to uncertainties in the
  relative calibration of both instruments. Specifically, the
  NIRSpec/G395H
spectrum, where the CO band lies, was multiplied by the quoted factor 
to match the continuum of MIRI/MRS CH1-short in the
$4.8-5.2$\,$\mu$m overlap region of the two detectors
(Fig.~\ref{conirspecmiri}a).

The lines in common in this overlap region allows us to cross-calibrate
the CO and H$_2$O bands. As shown in Fig.~\ref{conirspecmiri}a, most of
the CO P-branch lines within NIRSpec/G395H appear to be stronger
than within MIRI/MRS, after applying the scaling factor.
We compare in Fig.~\ref{conirspecmiri}b the two continuum-subtracted spectra,
where the NIRSpec spectrum has not been scaled up, and panel c
compares the corresponding CO P(16)-P(28) line absorption fluxes
from multi-Gaussian fit to the spectra.
From this comparison, the lines within NIRSpec/G395H
have not been scaled up by the 1.2 factor.

\section{The baseline of the CO band}
\label{basel}

\begin{figure*}
   \centering
\includegraphics[width=15cm]{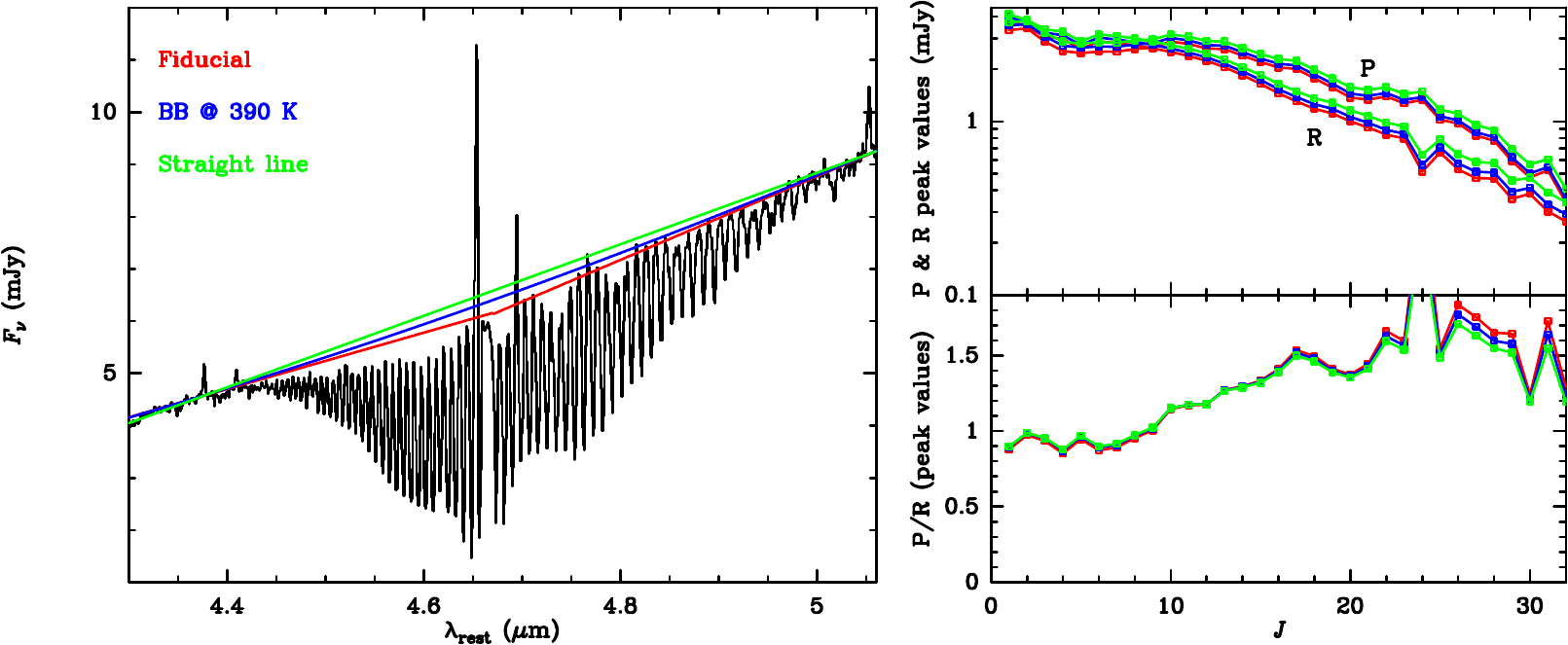}
\caption[]{{\it Left}: Comparison of considered baselines for the CO band.
  The red curve is the fiducial baseline we have used in this work.
  {\it Right-upper}: Peak absorption values of the CO P(J) and R(J)
  lines for the three baselines.
  {\it Right-lower}: The CO P-R asymmetry of the peak absorption 
  values for the 3 baselines.
}
\label{cobasel}%
\end{figure*}

The baseline continuum of the CO fundamental band is uncertain mostly
across the R-branch, because the lines are broad and blend at their wings
forming a pseudo/continuum. A similar effect is seen in the P-branch,
because of the contribution by the $^{13}$CO and C$^{18}$O
lines to the spectrum. Three alternative baselines are shown
in Fig.~\ref{cobasel}{\it left}: the fiducial one in red, which is
used in this work, a blackbody at 390\,K in blue, and a
straight line joining two extreme points of the band, in green.
The fiducial one is made up of 2 straight lines with different
slopes, joining at $\lambda_{\mathrm{rest}}=4.67$\,$\mu$m. 

The peak absorption and P/R values obtained with the use of
the 3 baselines are compared in Fig.~\ref{cobasel}{\it right}.
Comparing values using the fiducial and the straight line
baselines, the peak absorption increases by $10-15$\% across
the P-branch, except for the P(29) and P(30) that show an
increase of $18-20$\%.
Across the R-branch the peak values also increase similarly up
to $J=20$, and show a higher increase of $20-30$\% for
R($J>24$). Therefore, the P/R values remain similar up
to $J=20$, and decrease for the highest $J$ slightly as
compared with the errorbars shown in Fig.~\ref{peakabs}b.

Basically, the use of the straight line as the baseline would
increase the area covered by CO in all components of
Table~\ref{tab} by $10-15$\%. We mostly rely in the fiducial
baseline because our models for the CO band predict
that, across the P-branch, the continuum level is traced
between P(4)-P(5), P(10)-P(11)-P(12)-P(13), and
P(16)-P(17)-P(18)-P(19) (Fig.~\ref{bands}).

\section{Details on the H$_2$O band}
\label{h2olines}

\begin{figure*}
   \centering
\includegraphics[width=15.0cm]{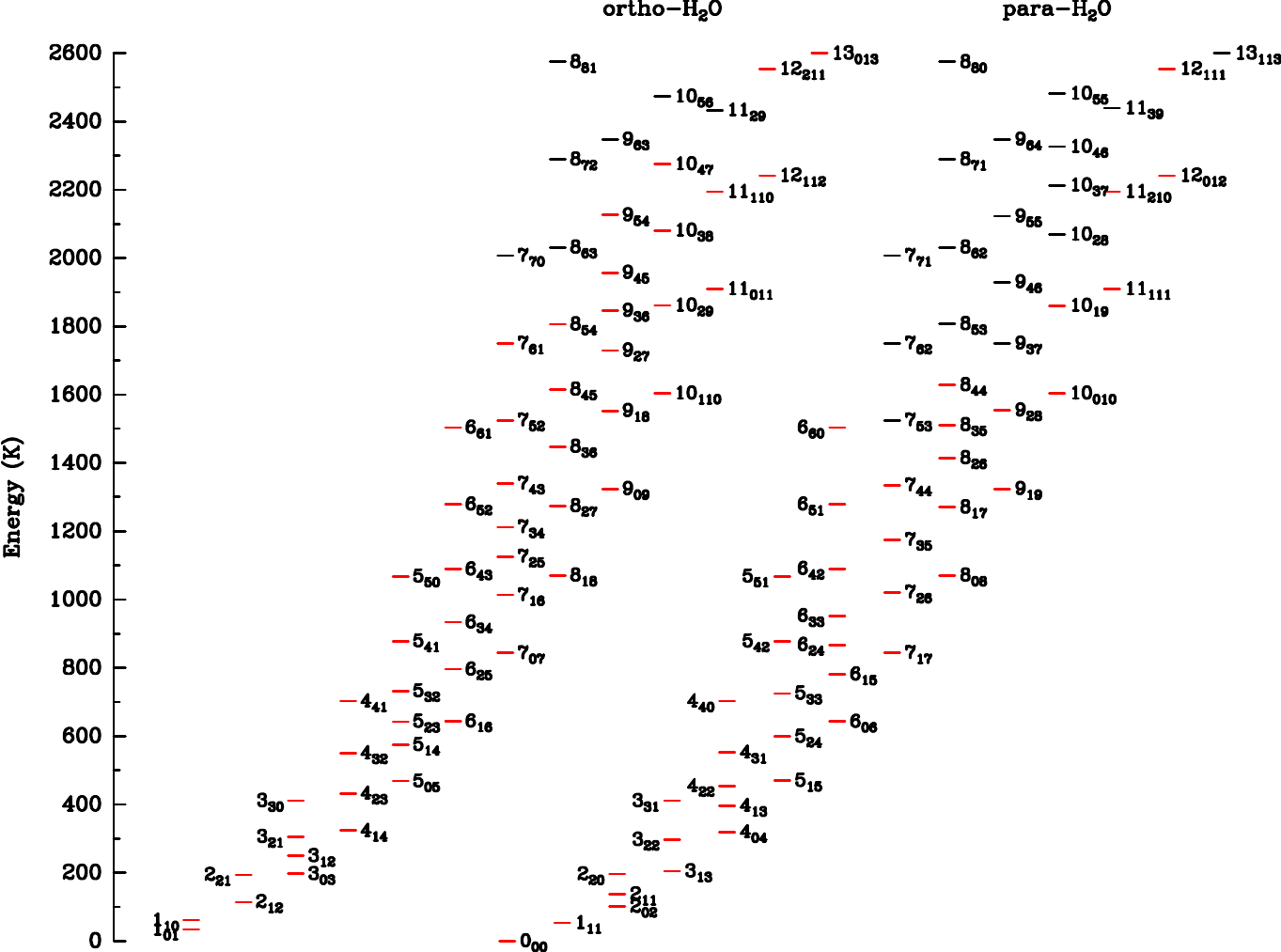}
\caption[]{Rotational energy level diagram of the H$_2$O $v=0$ vibrational
  state showing in red the levels from which at least one absorption line
  is detected in VV 114 SW-s2.
}
\label{h2oenerlev}%
\end{figure*}

Figure~\ref{h2oenerlev} shows in red all H$_2$O rotational levels of the ground
vibrational state from which at least one ro-vibrational line contributes,
according to our best-fit model, to the observed mid-IR spectrum of
VV~114 SW-s2.
As usual, the diagram is described by means of a series of $J$-ladders. 
It is well known that collisional excitation of H$_2$O, when thermal equilibrium
breaks down, usually tends to overpopulate the backbone rotational levels
(i.e. levels with $J=K_c$) at the expense to all others, due to the relatively
high optical depths and thus radiative trapping among the rotational lines
connecting these levels \citep[e.g.][]{jon73}.
Due to the high $A_{\mathrm{ul}}-$Einstein coefficients of the H$_2$O rotational
transitions, this will happen as long as
$n(\mathrm{H_2})\lesssim10^9$\,cm$^{-3}$,
depending on the column density per unit velocity interval.
In VV~114 SW-s2, however, ro-vibrational lines arising from levels
distant from backbone are detected (e.g. lines from all $J=6$ levels),
as well as ro-vibrational transitions from backbone levels up to
$J=13$. This illustrates the extreme excitation of H$_2$O in the $H_C$, and the
very high densities and column densities that would be required to
explain the observed spectrum via collisional excitation.

\begin{figure*}
   \centering
\includegraphics[width=16.0cm]{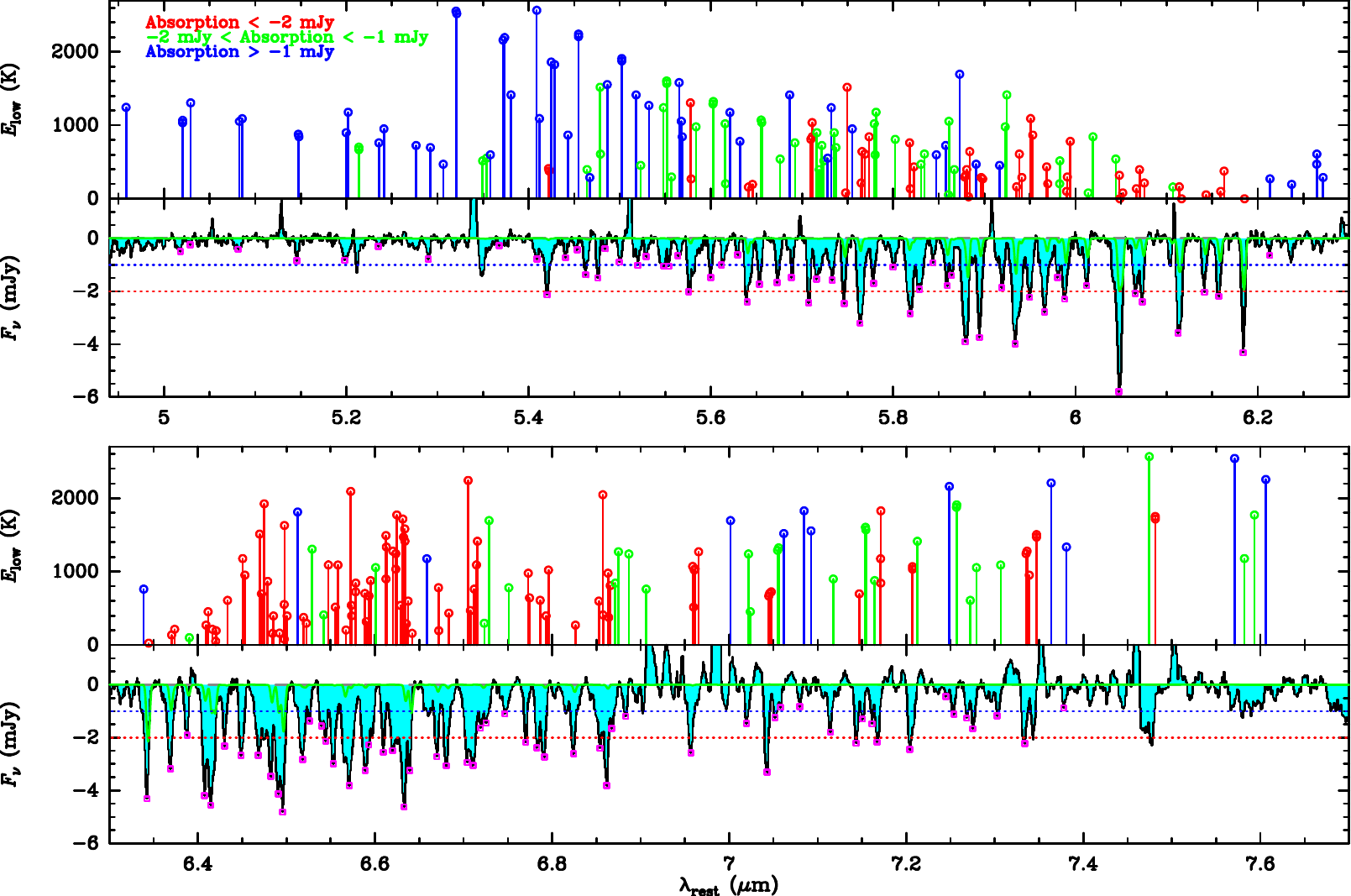}
\caption[]{The H$_2$O $\nu_2=1-0$ band lines that contribute significantly
  to the features in the $5.0-7.7$\,$\mu$m spectrum of VV 114 SW-s2.
  In the upper panels, the  $E_{\mathrm{low}}$ values are plotted
  with vertical segments ending in circles,
  and are coloured according to the absorption strength of the spectral
  feature they belong to relative to the horizontal dotted lines
  in the lower panels.
  The magenta squares overplotted on the spectrum indicate the peak flux
  values for the 116 spectral features that are used to compare the
  band with model results (Fig.~\ref{peakabs}c-d).
}
\label{h2obandelow}%
\end{figure*}

The H$_2$O $\nu_2=1-0$ band spectrum is shown in Fig.~\ref{h2obandelow}.
The observed spectral features are marked with the lower level energy
($E_{\mathrm{low}}$) of the contributing transitions. 
Particularly crowded is the $6.4-6.8$\,$\mu$m range, with more than 70
transitions. Relatively strong features (absorption deeper than $-2$\,mJy,
marked in red in the upper panels) usually include low-excitation lines
($E_{\mathrm{low}}\lesssim500$\,K) but some high-excitation transitions
($E_{\mathrm{low}}\gtrsim800$\,K) also generate strong absorption
($8_{18}-7_{07}$, $8_{08}-7_{17}$, and $8_{27}-8_{18}$ at $5.71$\,$\mu$m;
$7_{34}-7_{25}$ and $6_{33}-6_{24}$ at $5.95$\,$\mu$m;
$7_{25}-7_{34}$ and $6_{24}-6_{33}$ at $6.45$\,$\mu$m;
$7_{16}-7_{25}$ at $6.55$\,$\mu$m;
$7_{43}-7_{52}$, $6_{25}-6_{34}$, and  $7_{35}-7_{44}$ at $6.61$\,$\mu$m;
$4_{32}-5_{41}$, $6_{24}-7_{35}$, and  $9_{18}-10_{29}$ at $7.17$\,$\mu$m;
$4_{41}-5_{50}$ and $4_{40}-5_{51}$ at $7.21$\,$\mu$m;
$5_{41}-6_{52}$, $5_{42}-6_{51}$, and  $5_{24}-6_{33}$ at $7.34$\,$\mu$m;
$5_{50}-6_{61}$ and $5_{51}-6_{60}$ at $7.35$\,$\mu$m;
$6_{52}-7_{61}$ and $6_{51}-7_{62}$ at $7.48$\,$\mu$m).
This illustrates the extraordinary excitation conditions in the $H_C$.

  To explain this excitation via radiative pumping rather than collisions,
  the source of mid-IR radiation should be optically thick in the mid-IR;
  otherwise, $T_{\mathrm{rot}}$ within the H$_2$O ground vibrational 
  state would decrease below the required values.
  Then, the column density of neutral gas associated with the blackbody
  emission is $N_{\mathrm{H}}\gtrsim10^{23.5}$\,cm$^{-2}$ 
  (i.e. $\tau_{\mathrm{6\mu m}}\gtrsim3$) on $0.3$\,pc scales, giving a
  lower limit on the density of $n_{\mathrm{H}}\gtrsim10^{5.5}$\,cm$^{-3}$.
  Trapping of high-energy photons within this compact region will make
  the source dim in the X-ray bands \citep{ina16}. Fine-structure lines
  of high-ionization species such as [\ion{Ne}{v}]\,14.32\,$\mu$m
   and [\ion{O}{iv}]\,25.89\,$\mu$m \citep[e.g.,][]{per10} are not detected 
  in SW-s2, but are expected to be strongly quenched by
  this trapping, mid-IR extinction, and possibly by
  collisional de-excitation.

Our fiducial model in Fig.~\ref{bands} captures
the general behaviour of the H$_2$O band. The contribution of
the $W_C$ (Fig.~\ref{peakabs}c-d) is important for the low excitation
lines lying beween $5.6$ and $6.8$\,$\mu$m, but negligible on both
ends of the band. We note that the MRS spectrum, with high
quality for $\lambda_{\mathrm{rest}}<6.9$\,$\mu$m, shows ripples 
at longer wavelengths where the measured peak absorption of
the H$_2$O spectral features have thus larger uncertainties
(Fig.~\ref{peakabs}d).
Across the H$_2$O band (Fig.~\ref{bands}),
there are still some spectral features that are unidentified
or significantly underestimated. Among the former, the $6.313$ and
$6.325$\,$\mu$m adjacent features, the $6.660$\,$\mu$m absorption
in between the $7_{26}-7_{35}$ and $1_{11}-2_{20}$ lines, and the
shoulders flanking the $2_{12}-3_{21}$ line at $6.826$\,$\mu$m.
The spectral features that are underestimated by more than
35\% are:
$5.321$\,$\mu$m ($13_{112}-12_{211}$ \& $13_{212}-12_{111}$),
$5.440$\,$\mu$m ($7_{25}-6_{24}$),
$6.237$\,$\mu$m ($3_{13}-2_{20}$),
$6.265$\,$\mu$m ($6_{16}-5_{23}$),
and $7.378$\,$\mu$m ($6_{33}-7_{44}$) (but the strength of the
latter feature is relatively uncertain).

As indicated in the main text, the areas $A_{\mathrm{CO,H_2O}}$ 
are hard to constrain, specifically because of their
dependence on the thickness of the absorbing shell and on the
velocity field.
There is a mismatch of $\sim30$\% between $A_{\mathrm{CO}}$ and
$A_{\mathrm{H_2O}}$ in the $H_C$ (and thus between the predicted associated
continua in Fig.~\ref{maps}g) that may be due to these dependences.
The H$_2$O-to-CO abundance ratio in the $H_C$ is $\sim1$, 
orders of magnitude higher than the typical values derived in cold
molecular gas \citep[e.g.][]{mel20}. This is in line with expectations for
hot (several hundred K) material, as in such environments icy grain
mantles will have been vaporized and any atomic oxygen in the gas phase
will have been converted to H$_2$O in high temperature neutral-neutral
reactions with H$_2$. The mismatch in areas is also seen in
the $W_C$, but in this case can be attributed to the different excitation
of both species in more moderate environments, where CO will be more
widespread than H$_2$O.

  The H$_2$O $\nu_1=1-0$ and $\nu_3=1-0$ bands lie at $2.5-3.0$\,$\mu$m,
  and the $\nu_3$ band is the strongest. This $\nu_3$
  band is not detected in SW-s2. We have generated LTE models for it,
  using the same parameters as in our LTE models for the $\nu_2$
  band in Fig.~\ref{peakabs}c-d but with a background radiation
  temperature of $T_{\mathrm{bck}}=1400$\,K (Fig.~\ref{maps}g).
  If the absorbing H$_2$O gas were fully covering the $2.5-3.0$\,$\mu$m
  continuum as well, the models indicate that strong absorption would
  be detected in the H$_2$O $\nu_3$ band; the lack of detection indicates that
  H$_2$O is covering $\lesssim5$\% of the $2.5-3.0$\,$\mu$m continuum
  emission. Therefore, the nir continuum from SW-s2 is not
  related to the $4.4-8.0$\,$\mu$m continuum associated
  with the CO $v=1-0$ and H$_2$O $\nu_2=1-0$ bands.
  This result is fully consistent with our model
  for the continuum in Fig.~\ref{maps}g, involving strong
  attenuation of the component that dominates the
  mid-IR $4.4-8.0$\,$\mu$m emission 
   (see also Appendix~\ref{extinc}).

\section{The mid-IR extinction}
\label{extinc}

\begin{figure}
   \centering
\includegraphics[width=8.0cm]{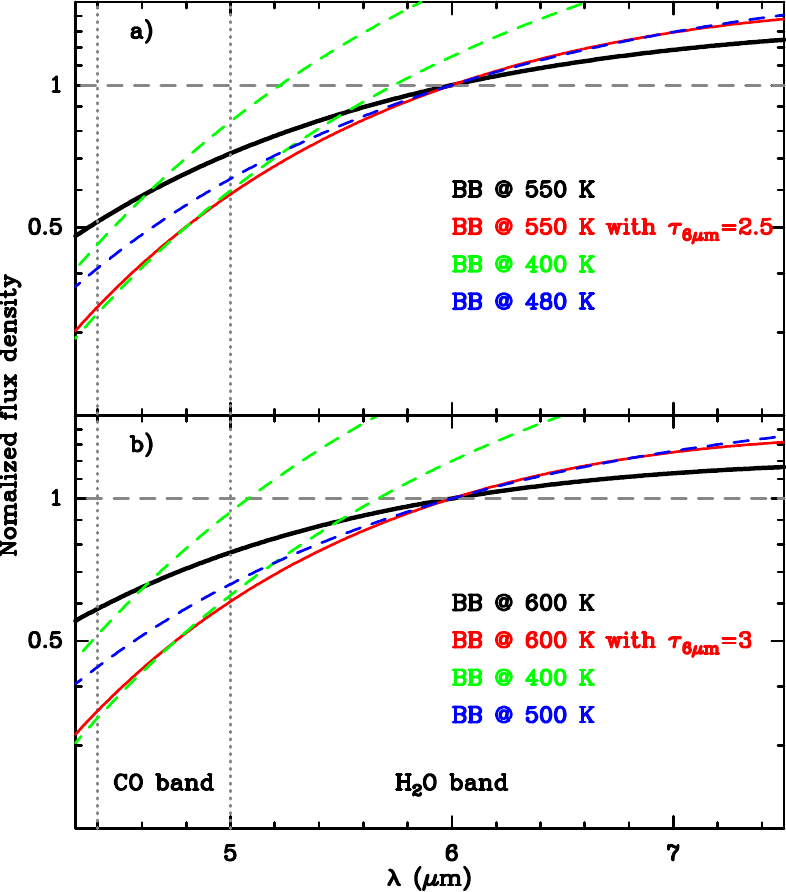}
\caption[]{Effect of extinction on the mid-IR continuum slope.
  The broad black lines show the shape of a blackbody source
  with {\bf a)} $T=550$\,K, and {\bf b)} $T=600$\,K. The red
  curves indicate how the SED changes when the blackbody
  emission is affected by extinction with
  $\tau_{\mathrm{6\mu m}}^{\mathrm{ext}}=2.5-3$, using the
  extinction law derived by \cite{ind05} and \cite{chi06}.
  The dashed green and blue curves are reference blackbodies
  at 400 and $480-500$\,K.
}
\label{vv114extinc}%
\end{figure}

We show in Fig.~\ref{vv114extinc} the extinction required to
match the $T_{\mathrm{app}}$ values inferred in VV 114 SW-s2,
assuming intrinsic blackbody emission with $T=550$\,K (panel a)
and 600\,K (panel b). These blackbody curves have a slope
across the CO band ($4.4-5.0$\,$\mu$m) very different from
$T_{\mathrm{app}}\sim400$\,K (green upper curves), which is the
value inferred from the CO P-R asymmetry. After applying
the extinction law by \cite{ind05} and \cite{chi06}
with $\tau_{\mathrm{6\mu m}}^{\mathrm{ext}}=2.5-3$, the resulting
reddened slopes (red curves) at $4.4-5.0$\,$\mu$m approach 
the required steep values. Extinction also changes the slope
across the H$_2$O band, with $T_{\mathrm{app}}\sim500$\,K at long
wavelengths (blue curves). High values of
$\tau_{\mathrm{6\mu m}}^{\mathrm{ext}}=2.5-3$, which attenuate
the intrinsic continuum emission by $\sim1$\,dex, are needed
to match the $T_{\mathrm{app}}$ values derived from the molecular
bands.

\section{CO band profiles and the best-fit model}
\label{coprofiles}

\begin{landscape}
\begin{figure}
\centering
\includegraphics[width=1.25\textwidth, angle=0]{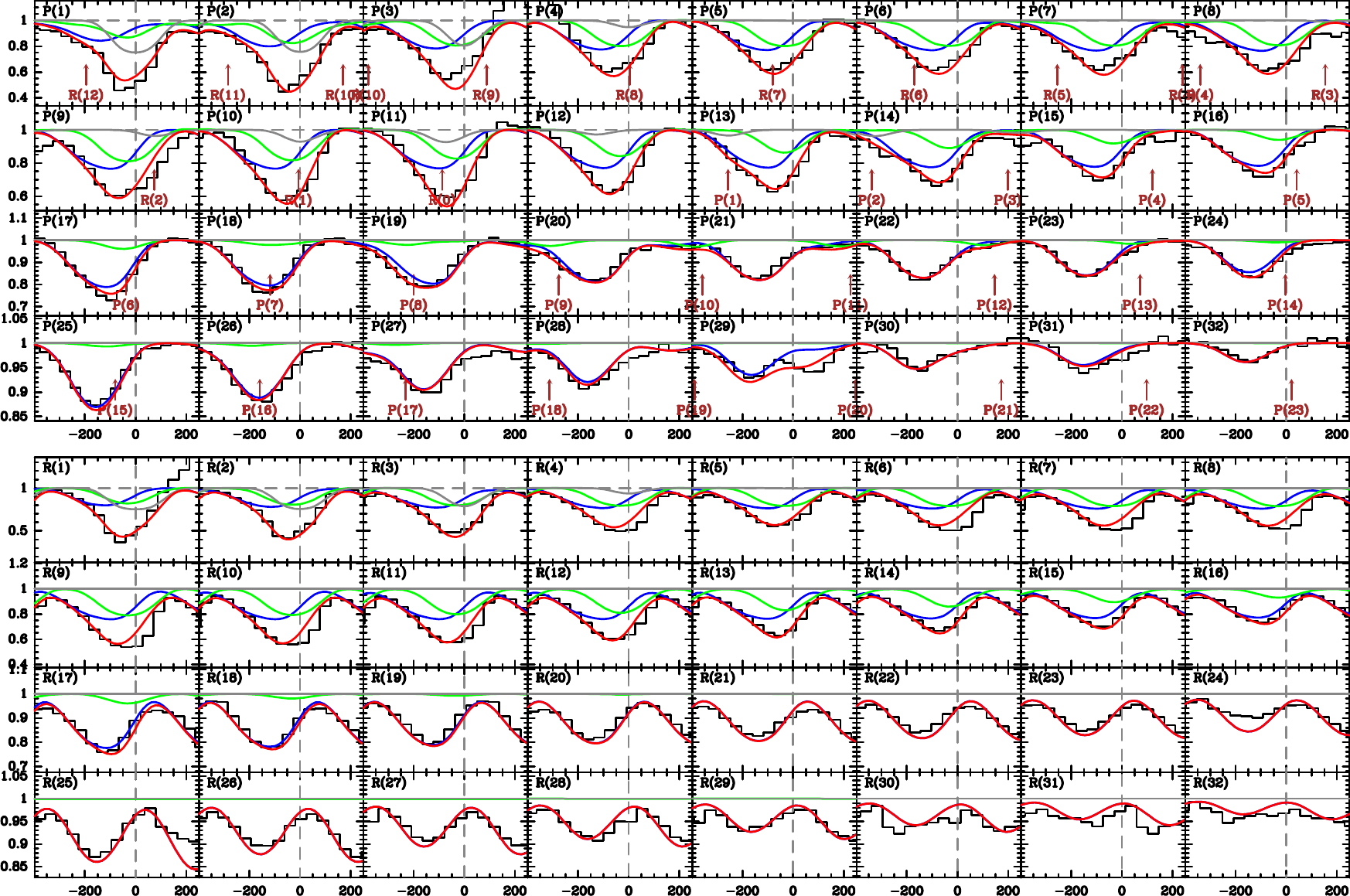}
\caption{Continuum-normalized profiles of the CO $v=1-0$ lines up to $J=32$.
  The positions of the $^{13}$CO lines are indicated in brown. 
  The best-fit model is overlaid in red, and the contribution by the
  individual components is also shown ($H_C$: blue, $W_C$: green, and
  $C_C$: gray). The abscissa axis is velocity in km\,s$^{-1}$ relative
  to the redshift inferred from the CO ($3-2$) line, $z=0.02013$.
}
\label{coprof}
\end{figure}
\end{landscape}

The continuum-normalized profiles of the CO $v=1-0$ lines up to $J=32$
are compared with predictions of our best-fit model in
Fig.~\ref{coprof}. The positions of the $^{13}$CO ro-vibrational lines
are indicated in brown.

The model fit is globally satisfatory, but there are some
significant discrepancies.
First, the CO P(1) line is not fully reproduced, probably indicating
the presence of additional cold gas along the line-of-sight to the
continuum source that the $C_C$ does not account for. A probably
related issue is that the absorption between CO P(8) and P(9)
and between CO P(14) and P(15) are not fully reproduced
(see also Fig.~\ref{bands}). These absorption are
coincident with the rest position of the $^{13}$CO R(3) and P(3) lines,
most likely indicating that the column density of the $C_C$
is underestimated. Finally, we note that the CO R(4)-R(13)
lines are all underpredicted at central and redshifted velocities,
but this effect is not seen in the P-branch counterpart.
This is indicative of the complexity of the $W_C$, which
most likely probes several components associated with the
mid-IR source, and some of them could contribute with different
absorption strengths in the two branches \citep{per23,gber23}.

\section{Estimating the BH mass from lifetime constraints}
\label{mbhderived}

Eddington luminosity ($L_{\mathrm{Edd}}$) and BH mass ($M_{\mathrm{BH}}$)
are equivalent through the standard relation $M_{\mathrm{BH}}= A\,L_{\mathrm{Edd}}$,
where $A=3.0\times10^{-5}\,M_{\odot}L_{\odot}^{-1}$.
However, the actual luminosity $L$,
\begin{equation}
  L\sim10^{10}\exp\{\tau_{\mathrm{6\mu m}}^{\mathrm{ext}}-2.5\}\,L_{\odot},
    \label{eq:lum}
\end{equation}
will generally depart from $L_{\mathrm{Edd}}$, and we write
\begin{equation}
  M_{\mathrm{BH}} = \frac{A}{\epsilon}\,L
  \label{eq:mbh}
\end{equation}
where $\epsilon\equiv \frac{L}{L_{\mathrm{Edd}}}$.
The mass accretion rate is given by
\begin{equation}
  \dot{M}_{\mathrm{BH}}  = \frac{L}{\eta c^2}
  \label{eq:mdot}
\end{equation}
where $\eta$ is the radiative efficiency. We can also define
$\eta_{\mathrm{Edd}}\equiv0.1$, but the actual value of $\eta$ may also
depart from $\eta_{\mathrm{Edd}}$.
Using eqs.~(\ref{eq:mbh}) and ~(\ref{eq:mdot}),
\begin{equation}
  \dot{M}_{\mathrm{BH}}  = \frac{\epsilon}{A\,c^2\eta }M_{\mathrm{BH}}
  \label{eq:mmdot}
\end{equation}
with the solution
\begin{equation}
  M_{\mathrm{BH}}(t)=M_{\mathrm{BH,0}}\exp\{t/t_0\},
  \label{eq:mmdotsol}
\end{equation}
where we have assumed that $\epsilon/\eta$ remains constant. The
$e-$folding time is
\begin{equation}
t_0= \frac{A\,c^2\eta }{\epsilon}
  \label{eq:t0}
\end{equation}

\begin{figure}
   \centering
\includegraphics[width=8.0cm]{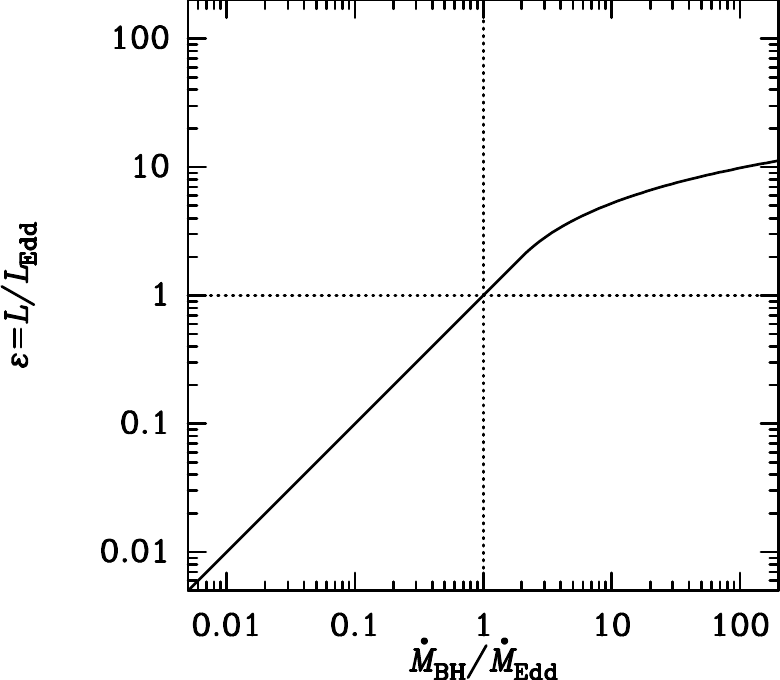}
\caption[]{
    Adopted relationship between $\epsilon\equiv L/L_{\mathrm{Edd}}$
  and the mass accretion rate onto the BH relative to the Eddington value;
  as derived by \cite{wat00}. Note that $\dot{M}_{\mathrm{Edd}}$ is here
  defined as $10\times$ the value defined in \cite{wat00}.
}
\label{epsilon}%
\end{figure}

To proceed further, we need $(i)$ the relationship between $\epsilon$ and
$\eta$, and $(ii)$ the time constraint for the BH growth:
\begin{itemize}
\item[$(i)$] The relationship derived by \cite{wat00} for the
  ``slim disk'' model is used, which we re-write here as in \cite{toyi21}:
\begin{equation}
  L = \begin{cases}
    2L_{\mathrm{Edd}}
  \left[ 1 + \ln\left(\frac{\dot{M}_{\mathrm{BH}}}{2\dot{M}_{\mathrm{Edd}}}\right)
    \right] & (\mathrm{if} \,\, \dot{M}_{\mathrm{BH}}>2\dot{M}_{\mathrm{Edd}})
  \\
  L_{\mathrm{Edd}}
  \left(\frac{\dot{M}_{\mathrm{BH}}}{\dot{M}_{\mathrm{Edd}}}\right) &
  \mathrm{(otherwise)}
  \end{cases}
\end{equation}
where $\dot{M}_{\mathrm{Edd}}\equiv L_{\mathrm{Edd}}/(\eta_{\mathrm{Edd}} \, c^2)$.
The curve is shown in Fig.~\ref{epsilon}.
$\epsilon\equiv \frac{L}{L_{\mathrm{Edd}}}$ flattens for
$\dot{M}_{\mathrm{BH}}/\dot{M}_{\mathrm{Edd}}>2$
accounting for the low radiative efficiency at high accretion rates, but
can attain values $\epsilon\sim10$. Similar excesses of $L$ over
$L_{\mathrm{Edd}}$ together with beaming effects are the basis for the
interpretation of most ultraluminous X-ray sources (ULXs) as super-Eddington
accreting stellar BHs or neutron stars in binary systems
\citep[XRBs, e.g.][]{beg06,pou07,kin08}.
The above relationship can be rewritten in terms of $\epsilon$ and $\eta$:
\begin{equation}
  \epsilon = \begin{cases}
2\left[ 1 + \ln\left(\frac{\epsilon \,\eta_{\mathrm{Edd}}}{2\eta}\right)\right] 
& \mathrm{(if  \,\, \epsilon \,\eta_{\mathrm{Edd}}>2\eta) } \\
\frac{\epsilon \,\eta_{\mathrm{Edd}}}{\eta} \rightarrow \eta=\eta_{\mathrm{Edd}}
& \mathrm{(otherwise)}
  \end{cases}
  \label{eq:eps}
\end{equation}
\item[$(ii)$] The fiducial time $t$ required to attain the current mass is
  (Section~\ref{mbh})
\begin{equation}
t \sim 10\,\mathrm{Myr},
  \label{eq:t}
\end{equation}
with an uncertainty of a factor $\sim2$. We estimate the $e-$folding
time $t_0$ using eq.~(\ref{eq:mbh}) and our result for the luminosity $L$ in
eq.~(\ref{eq:lum}), 
\begin{equation}
  M_{\mathrm{BH}} (t) =
  \frac{3\times10^{5}\,M_{\odot}}{\epsilon}
  \exp\{\tau_{\mathrm{6\mu m}}^{\mathrm{ext}}-2.5\} 
  \label{eq:mbhres0}
\end{equation}
Using now eq.~(\ref{eq:mmdotsol}), and assuming an initial BH seed with mass
$M_{\mathrm{BH,0}}\sim100\,M_{\odot}$, we obtain
\begin{equation}
  t=t_0\left[\ln\left(\frac{3\times10^{3}}{\epsilon}\right)+
  (\tau_{\mathrm{6\mu m}}^{\mathrm{ext}}-2.5)\right]
\end{equation}
We expect high accretion rates with $\epsilon$ in the range $[1,10]$,
and then
\begin{equation}
t=t_0\left[(5.7-8)+(\tau_{\mathrm{6\mu m}}^{\mathrm{ext}}-2.5)\right],
\end{equation}
where the values $5.7$ and $8$ correspond to $\epsilon=10$ and 1, respectively.
In view of this result and for simplicity, we can write the time constraint
in eq.~(\ref{eq:t}) in terms of the $e-$folding time as
\begin{equation}
t_0 = \frac{Ac^2\eta}{\epsilon} \sim 2^{+2}_{-1}\,\mathrm{Myr}.
  \label{eq:t0cons}
\end{equation}
\end{itemize}

With these prescriptions, we first note that the second solution of
eq.~(\ref{eq:eps}) and the time constraint eq.~(\ref{eq:t0cons})
are not compatible. Since in this case $\eta=\eta_{\mathrm{Edd}}=0.1$,
eq.~(\ref{eq:t0cons}) gives
\begin{equation}
  \epsilon = \frac{Ac^2\eta_{\mathrm{Edd}}}{t_{0}}=
  45.5\left(\frac{t_{0}}{1\,\mathrm{Myr}}\right)^{-1}
\end{equation}
where we have used that $A\,c^2=455$\,Myr. However,
this solution is only valid for
$\epsilon \,\eta_{\mathrm{Edd}}<2\eta$, that is, for $\epsilon<2$
($t_0\ge23$\,Myr).

We then resort to the first solution of eq.~(\ref{eq:eps}) and
(\ref{eq:t0cons}), which give 
\begin{equation}
  t_{0} = \frac{Ac^2\eta_{\mathrm{Edd}}}{2\exp\{\epsilon/2-1\}},
\end{equation}
which is used to evaluate $\epsilon$:
\begin{equation}
  \epsilon =
  2\left[\ln\left(\frac{Ac^2\eta_{\mathrm{Edd}}}{2t_{0}}\right)+1\right]
  = 8.2-2\ln t_{0},
\end{equation}
where $t_{0}$ is in Myr. Inserting this constraint in eqs.~(\ref{eq:lum}) and
(\ref{eq:mbh}), we obtain an estimate of the mass of the BH:
\begin{equation}
  M_{\mathrm{BH}} = \frac{A}{\epsilon}\,L \sim
    \frac{3.7\times10^4\,M_{\odot}}{1-0.244 \ln t_{0}} \,
  \exp\{\tau_{\mathrm{6\mu m}}^{\mathrm{ext}}-2.5\},
\end{equation}
which is an IMBH as long as $t_{0}<15$\,Myr. For the fiducial $t_0\sim2$\,Myr,
$M_{\mathrm{BH}}\sim 4.5\times10^4\,M_{\odot}$, and the current mass accretion rate
is then
\begin{equation}
  \dot{M}_{\mathrm{BH}}  = \frac{M_{\mathrm{BH}}}{t_{0}} \sim
  2.3\times10^{-2} \exp\{\tau_{\mathrm{6\mu m}}^{\mathrm{ext}}-2.5\} \,\,
  M_{\mathrm{\odot}}\, \mathrm{yr^{-1}},
\end{equation}
with $\eta\sim2.8\times10^{-2}$ and
$\dot{M}_{\mathrm{BH}}/\dot{M}_{\mathrm{Edd}}\sim25$.

\end{appendix}

\end{document}